\documentclass[a4paper, 11pt]{article}
\usepackage[a4paper,top=3cm,bottom=3cm,left=3cm,right=3cm]{geometry}
\usepackage{booktabs}
\usepackage{amsmath}
\usepackage{color}
\usepackage{graphics}
\usepackage{graphicx}
\usepackage{enumitem}
\usepackage{wrapfig}
\usepackage{amssymb}
\usepackage{hyperref}
\usepackage[utf8]{inputenc}
\usepackage{amsfonts}
\usepackage[colorinlistoftodos]{todonotes}
\usepackage{algpseudocode}
\usepackage{cite}

\DeclareMathOperator*{\imunit}{\mathrm{i}}

\definecolor{myblue}{RGB}{0,91,150}
\definecolor{mylightgrey}{RGB}{234,234,236}
\definecolor{mygrey}{RGB}{156,165,164}
\definecolor{mygreenblue}{RGB}{84,178,169}
\definecolor{mylightblue}{RGB}{100,151,177}

\hypersetup{
	colorlinks=true,
	linkcolor=myblue,
	filecolor=mygrey,      
	urlcolor=mygreenblue,
	citecolor=mygreenblue,
}

\title{\LARGE{\textbf{Glassy dynamics on networks:\\local spectra and return probabilities}}}
\author{Riccardo Giuseppe Margiotta$^{\mathrm{1, *}}$, Reimer K\"uhn$^{\mathrm{1}}$, Peter Sollich$^{\mathrm{1,2}}$\\[0.2cm]${}^1$ King's College London, Department of Mathematics, Strand,\\London WC2R
	2LS, United Kingdom
\\
${}^2$ Institute for Theoretical Physics, Georg-August-University G\"ottingen\\
Friedrich-Hund-Platz 1, D-37075 G\"ottingen, Germany
	\\[0.2cm]$^*$\emph{riccardo\_giuseppe.margiotta@kcl.ac.uk}}

\begin{document}
\maketitle


\begin{abstract}
The slow relaxation and aging of glassy systems can be modelled as a Markov process on a simplified rough energy landscape: energy minima where the system tends to get trapped are taken as nodes of a random network, and the dynamics are governed by the transition rates among these. In this work we consider the case of purely activated dynamics, where the transition rates only depend on the depth of the departing trap. The random connectivity and the disorder in the trap depths make it impossible to solve the model analytically, so we base our analysis on the spectrum of eigenvalues $\lambda$ of the master operator. We compute the local density of states $\rho(\lambda|\tau)$ for traps with a fixed lifetime $\tau$ by means of the cavity method. This exhibits a power law behaviour $\rho(\lambda|\tau)\sim\tau|\lambda|^T$ in the regime of small relaxation rates $|\lambda|$, which we rationalize using a simple analytical approximation. In the time domain, we find that the probabilities of return to a starting node have a power law-tail that is determined by the distribution of excursion times $F(t)\sim t^{-(T+1)}$. 
We show that these results arise only by the {\em combination} of finite configuration space connectivity and glassy disorder, and interpret them in a simple physical picture dominated by jumps to deep neighbouring traps.
\end{abstract}


\section{Introduction}
In pursuing a better understanding of non-equilibrium glassy systems, scientists have invested much effort into characterising their complex, multidimensional potential energy landscapes in configuration space. Key properties of these energy landscapes are the number of minima, the distribution of their depths, and of the heights of barriers between them. These have been explored using both computer simulations \cite{Buchner1999, Heuer2008, DeSouza2009} and theoretical approaches \cite{Bray1980,Cavagna1998,Fyodorov2012}. The picture that has emerged is that the energy landscape of glasses is extremely complex, consisting of (exponentially many) minima, barriers and saddles of any order. The crystalline configurations that would be occupied in equilibrium at low temperature are  hidden in this maze of valleys and walls and this keeps glassy systems out of equilibrium on typical observation timescales. When a glass is prepared, for example by quenching a viscous liquid to a low enough temperature, the system is expected to start descending rapidly towards a local minimum of the energy landscape \cite{Biroli2001, Berthier2011} (though how it does so is in itself not trivial \cite{Chacko2019}).  As time proceeds, the system will then slowly explore progressively lower energy minima, leading to aging effects where physical properties depend on the time since preparation of the glass \cite{Bouchaud_Cugliandolo1997}. The dynamics in this regime can be thought of as consisting mostly of thermal fluctuations around an energy minimum, interspersed with rare large fluctuations that allow the system to cross a barrier and reach a new energy minimum. If one ignores the small thermal fluctuations around a given minimum, i.e.\ within a given ``basin'', and focusses on the  long time exploration of the various basins, then glassy dynamics can be modelled as a Markov process on a network, with each local energy minimum represented as a network node. A complete definition of such a model requires assumptions on the network topology, i.e.\ the connectivity in configuration space, and the transition rates between the nodes. The latter are expressed in terms of the energies of the various energy minima and the barriers between them. The distribution of the energy minima that enters here is expected on  general grounds to have an exponential tail towards the deepest minima~\cite{Odagaki1995,Bouchaud1997}.

The trap model \cite{Bouchaud1992, Monthus1996, Bovier2005} is one of the most successful descriptions of glassy dynamics that belong to this framework, where the energy minima are thought of as traps ``hanging off'' a threshold level where all barriers are located; in addition, the network of traps is assumed to be fully connected so that every trap can be accessed from any other. All moves between traps then require activation to the threshold level, which is convenient to use as the zero of the energy scale, and each jump takes the system to a randomly chosen new trap. The system thus effectively forgets with each jump what trap it was in before, making the dynamics a renewal process. The transition rates only depend on the departing energy depth because activation is always to the threshold level, and directly define the inverse lifetime of any trap. These simplifications allow the model to be solved analytically and give direct access to the evaluation of time dependent quantities. In particular, aging is described by two-time correlation functions that can be found explicitly and are given by the so-called arcsine law below the glass transition temperature \cite{BenArous2006b}. A variety of disordered and more complex models of glasses exhibit an emergent trap-like phenomenology and aging behaviour, as demonstrated by numerical evidence as well as analytical arguments \cite{Gayrard2016, Cammarota2018}.  However, the presence of dynamical correlations can make it hard to access the relevant timescales via simulations, and coarse-graining the evolution into larger effective basins may be required \cite{BaityJesi2018}. Importantly for us, the network of traps is generically {\em not} fully connected, and the original trap model then describes only the motion between the deepest effective basins at very long times. Note that a long time reduction of correlation functions to the arcsine law has been proven explicitly for the case of regular connectivity among the traps \cite{Ben_Arous2006}. Various works have investigated in particular the case of lattices~\cite{Monthus1996,Rinn2001,BenArous2006}, though this is more plausible when the dynamics is interpreted as describing movement of a particle in real space rather than of a system in a high-dimensional configuration space. In the latter case, disorder in the connectivity among traps \cite{Doye2002} inspired models of glassy dynamics on random networks. These are impossible to solve analytically, because of the disorder in trap depths and the random connectivity among nodes. Previous studies therefore had to rely on a heterogeneous mean field approximation \cite{Baronchelli2009,Moretti2011}, which is uncontrolled. A different approach can be taken, however, by basing the analysis on the spectral properties of the master operator. This operator is the continuous-time analogue of a Markov transition matrix, and is key in determining the dynamics of the system. In particular the spectral density or density of states (DOS) $\rho(\lambda)$ gives the spectrum of {\rm relaxation rates} of the system, and the localization properties of the eigenmodes carry information about the probability flow across the network. This is the approach that we followed in our previous work \cite{Margiotta2018}, which was dedicated to the analysis of trap models on sparse networks. In these models the zero energy threshold level remains present but jumps among traps are only allowed along network edges, i.e.\ local with respect to the network, so that the renewal property is lost.

In~\cite{Margiotta2018} we investigated the thermodynamic limit of an infinite network of traps by means of the cavity method. This approach exploits the local tree-like structure of networks with sparse connectivity and follows in this analogous applications to the spectral analysis of symmetric random matrices; see e.g. \cite{Rogers2008, Rogers2009, Metz2010, Kuhn2015, Benedetti2018} or \cite{Bordenave2010, Khorunzhy2004} for a rigorous discussion. We used two relevant limits as benchmarks: the mean field (MF) limit where the average connectivity diverges in the thermodynamic limit of infinite system size, thus giving the original Bouchaud trap model, and the infinite temperature or random walk (RW) limit, where the energy landscape no longer plays a role. Our findings confirmed the idea that the very long time dynamics is well described by the original fully connected trap model and does not depend on the topology of the network of traps: the DOS always has a small-$|\lambda|$ tail -- governing the long time relaxation -- with the same power law behaviour as in mean field, $\rho(\lambda)\sim|\lambda|^{T-1}$. This can be rationalized within a simple high temperature approximation. In addition to this, our results indicated a decomposition of the dynamics into three different timescales: the long time (small $|\lambda|$), network independent regime with localized eigenmodes, the short time region where eigenmodes are delocalized and dominated by the network connectivity, and an intermediate regime where the DOS is as in mean field but the eigenmodes are delocalized nonetheless.

In this work we significantly extend our analysis of the trap model on sparse networks by looking at the local DOS, $\rho(\lambda|\tau)$, which gives the contribution to the (total) DOS from all traps with a fixed average lifetime $\tau$. The high $T$ approximation scheme again proves useful for deriving an analytical approximation for $\rho(\lambda|\tau)$ in the regime relevant for the long time dynamics: we find $\rho(\lambda|\tau)\sim\tau|\lambda|^T$ when $|\lambda|\ll1/\tau$. These results are then translated into the time domain to give estimates for the return probability $P_{\tau}(t)$ and the distribution $F(t)$ of excursion times,  i.e. the times required by the system to return to an initial trap, leading to $P_{\tau}(t)\sim t^{-(T+1)}$ for $t\gg\tau$ and $F(t)\sim t^{-(T+1)}$. Remarkably, it is the distribution of deep minima surrounding the initial trap that determines these power laws, and they arise as a combined effect of limited connectivity \emph{and} trap depth disorder: if only one of these features is present, the local DOS becomes concentrated around $-1/\tau$ implying an exponential decay of the return probability.

The paper is organised as follows: after defining the model in section \ref{section_set_up}, we present our cavity theory in section \ref{section_local_DOS}, describe the approximation scheme and sketch the result for the power law tail of the local DOS; the full derivation is left to appendix \ref{appendix_second_shell_approximation}. In section \ref{section_return_probability} and \ref{section_excursion_times} we focus on the behaviour in the time domain by analysing, respectively, the return probability and the excursion time distribution. Finally, we summarise and discuss our results in section \ref{section_conclusion}.


\section{Bouchaud trap model on networks}\label{section_set_up}
We follow the set-up of our previous work \cite{Margiotta2018}: the problem is defined by a continuous-time Markov process on a sparse network (or graph), whose nodes (or traps) represent the minima of the energy landscape where system gets trapped. These have a positive energy that represents the {\em depth} of the minimum with respect to the level zero of the energy landscape, and determines the expected lifetime of that state. Trap depths ($E>0$) are quenched random variables following the exponential distribution $\rho_{E}(E)=\theta(E) \text{exp}(-E)$. The master equation defining the Markov process is 
\begin{equation}\label{master_equation}
\partial_t \mathbf{p}(t) = \mathbf{M} \mathbf{p}(t)
\end{equation}
where $\mathbf{p}(t)=(p_1(t),\ldots,p_N(t))$ is the probability distribution describing the position of the system on the network. The elements of the master operator $\mathbf{M}$ are:
\begin{equation}\label{master_operator}
M_{ij} = c_{ij} r_{ij} \quad \mbox{for}~i\neq j\ ,\qquad
M_{ii} = - \sum_{j (\neq i)} M_{ji}
\end{equation}
where $r_{ij}$ are the Bouchaud transition rates $r_{ij}= e^{-\beta E_j}/c\equiv r_j$, $c$ is the average connectivity of the network, $\beta$ is the inverse temperature and $c_{ij}$ is $1$ if $i$ and $j$ are connected, and $0$ otherwise. It is useful to define the quantity $\tau_j = (cr_j)^{-1}=\text{exp}(\beta E_j)$, which sets the scale of the expected waiting time ($c/k_j)\tau_j$ to leave a node $j$; here $k_j$ is the degree of the node. The distribution of energies $E$ implies a distribution for $\tau$ given by
\begin{equation}\label{lifetime_distribution}
\rho_{\tau}(\tau) = T \tau^{-(T+1)}
\end{equation}
Note that $\langle \tau \rangle$ diverges for $T\leq 1$, signalling a low $T$ regime where the dynamics gets glassy. In this work we will mostly focus on the case where the network connectivity is that of a random regular graph (RRG), i.e.\ where every node is connected to $c$ random nodes, with $c\geq 3$ so that the fraction of nodes outside the giant connected component of the graph vanishes in the large $N$ limit \cite{Bollobas2001}. This case is the simplest and yet it exhibits the same key features as more complex network topologies. This is confirmed by the results shown at the end of section \ref{subsection_cavity_method}, where the cases of Erd\"os-R\'enyi and scale-free connectivities are discussed.

The assumption of a configuration space characterised by a sparse and random connectivity makes the master equation (\ref{master_equation}) impossible to solve analytically. We therefore take another route and focus on the spectral properties of $\mathbf{M}$, whose $\alpha^{\text{th}}$ eigenvalue, left and right eigenvectors we write respectively as $\lambda_{\alpha}, \mathbf{w}_{\alpha}$ and $\mathbf{u}_{\alpha}$. A formal solution to (\ref{master_equation}) is then given by
\begin{equation}\label{p_t}
\mathbf{p}(t) = \sum_{\alpha=0}^{N-1} e^{\lambda_{\alpha} t} (\mathbf{w}_{\alpha}, \mathbf{p}(0))\mathbf{u}_{\alpha}
\end{equation}
where $(\cdot,\cdot)$ denotes the scalar product between vectors. If the network is connected there is only one vanishing eigenvalue $\lambda_0=0$, and the corresponding right eigenvector represents the equilibrium distribution of the system: $\mathbf{p}_{\text{eq}}=\lim_{t\to\infty}\mathbf{p}(t)=\mathbf{u}_0$. All other eigenvalues have a negative real part and the contribution of the associated eigenvectors to $\mathbf{p}(t)$ is exponentially suppressed over time. We will often refer to the eigenvectors of $\mathbf{M}$ as the eigenmodes (or simply the modes) of the dynamics, so e.g. we could say that the long time behaviour of the system is governed by the slow modes, thus referring to the eigenvectors in the small $|\lambda|$ regime. The importance of the spectrum of eigenvalues for the dynamics is evident as it provides the distribution of \emph{relaxation rates} of the system. For this reason, a central quantity for our analysis is the (total) density of states (DOS), defined as
\begin{equation}\label{total_DOS}
\rho(\lambda) = \frac{1}{N} \sum_{\alpha=0}^{N-1} \delta(\lambda-\lambda_{\alpha})
\end{equation}
One could equivalently consider the spectrum of the relaxation rates $r_{\alpha}=-\lambda_{\alpha}$, which would just flip the sign of lambda. We stick to the convention in (\ref{total_DOS}) for consistency with our earlier work \cite{Margiotta2018}. There we discussed in some detail the features of the DOS of the trap model defined on sparse networks. In all cases that we considered the DOS showed a $|\lambda|\to0$ power-law tail with the same exponent as found in the case of mean field connectivity, and eigenvectors exhibiting a localization transition, from delocalized fast modes to localized slow modes. We measured the degree of localization in terms of the inverse participation ratio (IPR), using a formula proposed by Boll\'e et al~\cite{Metz2010} to detect localization transitions in symmetric random matrices. In order to investigate the thermodynamic limit we relied on the cavity method, and used a population dynamics algorithm to solve the associated cavity equations numerically. This method links the master operator to the inverse covariance matrix of a complex Gaussian distribution and therefore requires a symmetric matrix as input, which in our case is obtained from the similarity transformation
\begin{equation}
\mathbf{M}^{\text{s}} = \mathbf{P}_{\text{eq}}^{-1/2} \mathbf{M} \mathbf{P}_{\text{eq}}^{1/2}
\end{equation}
Here $\mathbf{P}_{\text{eq}}$ is a diagonal matrix with non-zero elements given by the equilibrium distribution: $(\mathbf{P}_{\text{eq}})_{ii}=(\mathbf{p}_{\text{eq}})_i$. This transformation preserves the eigenvalue spectrum, which is real as $\mathbf{M}^{\text{s}}$ is real and symmetric. Also, it does not affect the diagonal elements of the master operator, i.e. $(\mathbf{M}^{\text{s}})_{ii}=(\mathbf{M})_{ii}$, a fact that will turn out to be crucial for us. The eigenvectors $\mathbf{v}_{\alpha}$ of $\mathbf{M}^{\text{s}}$ are given by $\mathbf{v}_{\alpha} = \mathbf{P}_{\text{eq}}^{-1/2} \mathbf{u}_{\alpha}=\mathbf{P}_{\text{eq}}^{1/2} \mathbf{w}_{\alpha}$. These retain the same localization properties as the eigenvectors of the original system, except for finite size effects mostly appearing close to the ground state (see \cite{Margiotta2018}, appendix E). The symmetry of $\mathbf{M}^{\text{s}}$ is a consequence of the detailed balance condition that holds between the transition rates and the equilibrium Boltzmann distribution $\mathbf{p}_{\text{eq}}$ \cite{vankampen2007spp}, see also \cite{Kurchan2009} for a discussion in the context of Fokker-Plank evolution.
Using a standard identity from random matrix theory \cite{Edwards1976}, the DOS of $\mathbf{M}^{\text{s}}$ can be expressed as 
\begin{equation}\label{total_DOS_G}
\rho(\lambda) = \lim_{\varepsilon\to 0} \frac{1}{\pi N} \sum_{i=1}^{N} \mathrm{Im} \, G_{ii}(\lambda_{\varepsilon}) 
\end{equation}
in terms of the resolvent 
\begin{equation}
\mathbf{G}(\lambda_{\varepsilon}) =(\lambda_{\varepsilon}\mathbf{I}-\mathbf{M}^{\text{s}})^{-1}
\end{equation}
Here  $\mathbf{I}$ indicates the $N\times N$ identity matrix and we have used the abbreviation $ \lambda_{\varepsilon} = \lambda - \imunit\varepsilon$ with $\imunit$ the imaginary unit and $\varepsilon$ small and positive.
In going from (\ref{total_DOS}) to (\ref{total_DOS_G}) one replaces the delta functions in (\ref{total_DOS}) with Lorentzians of width $\varepsilon$; thus $\varepsilon$ sets the numerical resolution that we have on the $\lambda$-axis when we come to evaluate quantities of interest, using in our case specifically the population dynamics algorithm outlined below. 

In this work we use the cavity method to study the DOS in more detail. In particular we decompose it into a set of {\em local} DOSs, one for each node; these local DOSs are defined explicitly below. The analysis allows to probe the important effects of heterogeneity in the network, as generated by the landscape of trap depths $E_i$. It will also enable us to obtain insights into the dynamics directly in the time domain, as e.g. time-dependent probabilities of return to a certain trap can be easily computed using results for the local DOSs.


\section{Local DOS}\label{section_local_DOS}

The $i^{\text{th}}$ term appearing in the sum on the right hand side of equation (\ref{total_DOS_G}) is the contribution to the total DOS given by a single node. To make this explicit, we can write 
\begin{equation}
\rho(\lambda)= \frac{1}{N}\sum_{i=1}^N\rho(\lambda|i)
\label{rho_decomp}
\end{equation}
with $\rho(\lambda|i) = \lim_{\varepsilon\to 0} \mathrm{Im} \, G_{ii}(\lambda_{\varepsilon})/\pi$. This quantity is referred to as the (single node) \emph{local DOS}. By translating the eigendecomposition of $\mathbf{M}^{\rm s}$ into one for the resolvent matrix, $\mathbf{G}(\lambda_\varepsilon) = \sum_\alpha (\lambda_\varepsilon-\lambda_\alpha)^{-1} \mathbf{v}_\alpha\mathbf{v}_\alpha^{\rm T}$, one sees that the local DOS can be written more explicitly as
\begin{equation}\label{local_DOS_single_site}
\rho(\lambda|i)= \sum_{\alpha=0}^{N-1} \delta(\lambda-\lambda_{\alpha}) v_{\alpha,i}^2
\end{equation}
The normalization of eigenmodes implies that summing over all network nodes $i$ and dividing by $N$ gives the total DOS defined by (\ref{total_DOS}), as written in (\ref{rho_decomp}). More generally, it is possible to decompose the total DOS according to the contribution from all traps with a given property, e.g.\ a fixed local timescale $\tau_i=\tau$. Following this idea, we define $\rho(\lambda|\tau)$ as
\begin{equation}
\rho(\lambda|\tau)=\frac{1}{N_\tau}\sum_{i=1}^N \delta(\tau-\tau_i)\rho(\lambda|i)
\end{equation}
where $N_\tau=\sum_{i=1}^N \delta(\tau-\tau_i)$. The following relations then hold:
\begin{equation}\label{total_DOS_from_conditional_DOS}
\rho(\lambda) = \int \textrm{d}\tau \, \rho(\lambda|\tau)\rho_\tau(\tau)
\end{equation}
\begin{equation}
1=\int \textrm{d}\lambda \, \rho(\lambda|\tau)
\end{equation}
Here $\rho_\tau = N_\tau/N$ is the probability density function of $\tau$ for a given realization of the system with size $N$, which for $N\to\infty$ is self-averaging and given by the expression (\ref{lifetime_distribution}). Of course the same construction can be used to define a local DOS conditioned on generic local disorder variables; the node degree $k_i$ would be an obvious choice (see e.g. \cite{Kuhn2008}), though we do not pursue this here.

In the next section we show how to use the cavity method to evaluate the local DOS $\rho(\lambda|\tau)$, and we derive an analytical approximation valid for the small $|\lambda|$ tail based on the high $T$ approximation scheme that we introduced in \cite{Margiotta2018}.
%
%
\subsection{Cavity method}\label{subsection_cavity_method}
Here we only present a brief summary of the cavity construction and refer to our previous paper \cite{Margiotta2018} for a detailed derivation of the central result, i.e.\ the self-consistent equation for the distribution of cavity precisions given below. 

\begin{figure}[htb]
\centering
\includegraphics[width=6.75cm]{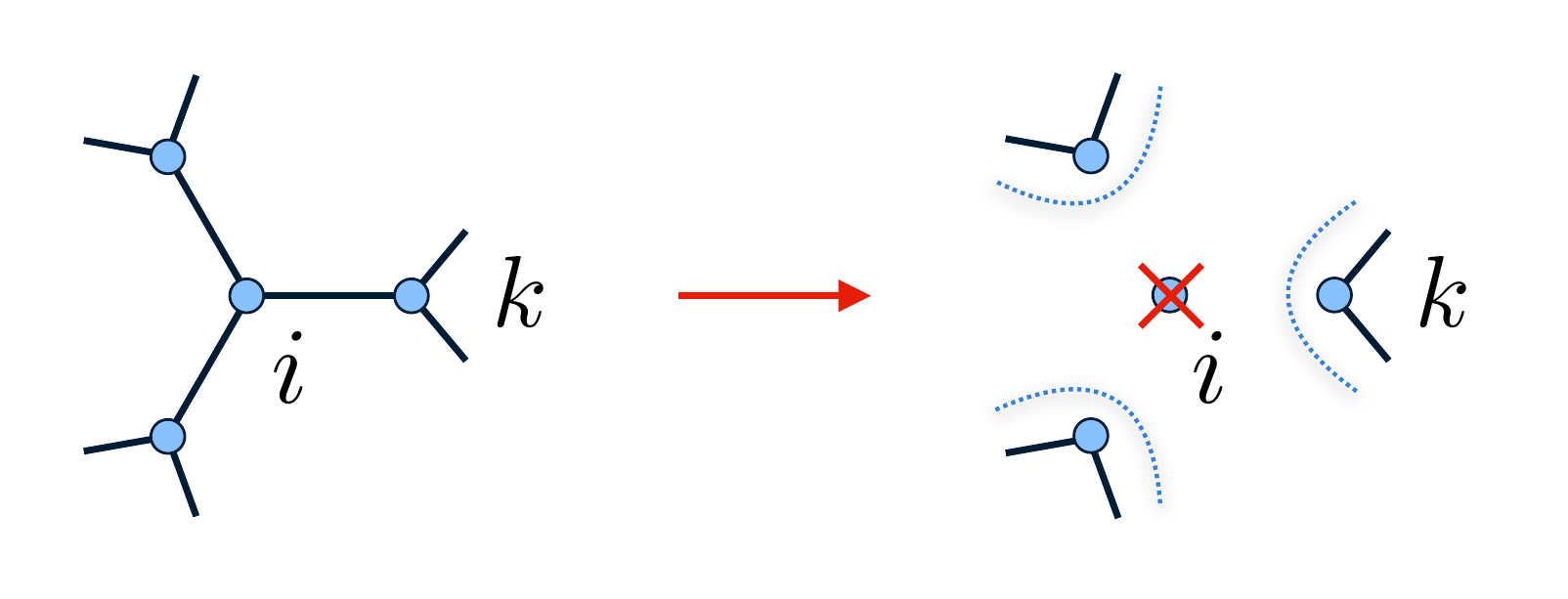}
	\caption{Local tree-like structure of a random regular graph with connectivity $c=3$ (left). When the central node $i$ is removed from the network (right), the branches become independent of each other in the thermodynamic limit $N\to\infty$, and so the joint probability distribution of the neighbourhood factorises: $P^{(i)}(\mathbf{y}_{\partial i})=\prod_{k\in\partial i}P^{(i)}(y_k)$.}\label{fig:cavity_sketch}
\end{figure}

First of all, one observes that the diagonal elements of the resolvent can be expressed as
\begin{equation}
G_{ii}= \imunit\int \mathrm{d}\mathbf{x}\, x_i^2 P(\mathbf{x}) = \imunit\int \mathrm{d} x_i\, x_i^2 P(x_i) 
\end{equation}
where $P(\mathbf{x})\propto \text{exp}(-\imunit \mathbf{x}^{\rm T} \mathbf{G}^{-1}\mathbf{x}/2)$ is a complex Gaussian measure with covariance matrix given by $\mathbf{G}$, and $P(x_i)$ the associated marginal distribution at node $i$. Exploiting the sparse structure of $\mathbf{G}^{-1}$, one can express the marginal distribution of $i$ in terms of the cavity distributions of its neighbouring nodes $k\in \partial_i$:
\begin{equation}\label{marginal_yi}
P(y_i) = e^{-\frac{\imunit}{2}\lambda_{\varepsilon}\frac{y_i^2}{r_i}}\prod_{k\in\partial_i}\int \mathrm{d} y_k e^{-\frac{\imunit}{2}(y_i-y_k)^2}P^{(i)}(y_k)
\end{equation}
Here $\partial_i$ indicates the neighbourhood of $i$, and we have used the change of variable $y_i=x_i \sqrt{r_i}$, which has the desired effect of confining the disorder from the transition rates $r_i$ to the diagonal terms. The last equation is based on the key assumption that the joint distribution of the nodes belonging to $\partial_i$ factorises when the central node $i$ is removed from the graph (see the illustration in Fig.~\ref{fig:cavity_sketch}). 
This is strictly true only when the network formed by the traps and allowed transitions between them is a tree, but it also provides a valid approximation whenever the topology of the network is at least locally tree-like, as is the case of sparse networks (e.g.\ with random regular or scale-free connectivity) in the large $N$ limit \cite{Albert2002}. Using the same line of reasoning as for (\ref{marginal_yi}) one can write for the cavity distributions the recursive relation
\begin{equation}\label{marginal_cavity_yk}
P^{(i)}(y_k) = e^{-\frac{\imunit}{2}\lambda_{\varepsilon}\frac{y_k^2}{r_k}}\prod_{l\in\partial_k \setminus i}\int \mathrm{d} y_l\,  e^{-\frac{\imunit}{2}(y_k-y_l)^2}P^{(k)}(y_l)
\end{equation}
Equation (\ref{marginal_cavity_yk}) is self-consistently solved by Gaussian distributions of the form $P^{(i)}(y_k)\propto \text{exp}(-\omega^{(i)}_k y_k^2/2)$. This ansatz transforms Eq.~(\ref{marginal_cavity_yk}) into an equivalent set of equations for the cavity precisions:
\begin{equation}\label{cavity_precisions}
\omega^{(i)}_k = \imunit \lambda_{\varepsilon} \tau_k c + \sum_{l\in\partial_k\setminus i} \frac{\imunit \omega_l^{(k)}}{\imunit+\omega_l^{(k)}}
\end{equation}
The Gaussian nature of the cavity marginals entails that single site marginals $P(y_i)$ are also Gaussian, 
with Eq.~(\ref{marginal_yi}) implying that single site precisions $\omega_i$ are of the form
\begin{equation}\label{precisions}
\omega_i = \imunit \lambda_{\varepsilon} \tau_i c + \sum_{k\in\partial_i} \frac{\imunit \omega_k^{(i)}}{\imunit+\omega_k^{(i)}},
\end{equation}
The system of equation for the cavity precisions (\ref{cavity_precisions}) can be solved recursively for a finite realization of the system. The marginal precisions $\omega_i$ are then obtained from (\ref{precisions}) and they give the diagonal elements of the resolvent via $G_{ii}=\imunit \tau_i c/ \omega_i$. 

We are concerned with the thermodynamic limit of a large network. Here we can exploit that for $N\to\infty$, where due to the locally tree-like assumption loops in the network become long, the different terms in the sum in (\ref{cavity_precisions}) become uncorrelated samples from the distribution $p(\omega)$ of cavity precisions. Requiring that the left hand side, too, is a sample from the distribution of cavity 
precisions, one obtains a self-consistency equation for $p(\omega)$. (Note that here and in the following we omit the superscript indicating the cavity graph in order to keep the notation simple.) For a general degree distribution $p_k$ this self-consistency equation reads
\begin{equation}\label{cavity_precisions_distribution}
p(\omega) = \sum_k \frac{k p_k}{c} \int \mathrm{d}\tau \rho_{\tau}(\tau)\prod_{l=1}^{k-1}\mathrm{d}\omega_l\,p(\omega_l)\, \delta(\omega - \Omega_{k-1})
\end{equation}
with the abbreviation
\begin{equation}\label{omega_function}
\Omega_{a} = \Omega_{a}(\lambda_{\varepsilon},\{\omega_l\},\tau) = \imunit\lambda_{\varepsilon}\tau c + \sum_{l=1}^{a} \frac{\imunit \omega_l}{\imunit + \omega_l}
\end{equation}
A numerical solution of (\ref{cavity_precisions_distribution}) can be found using a population dynamics algorithm; see \cite{Mezard2001} for a detailed explanation of this method. The core idea is to take an initialised population $\mathcal{P}$ of cavity precisions $\omega$ and then update each of them using the value of $\Omega_{k-1}$ given by a sample $\tau$ and $k-1$ random elements of $\mathcal{P}$ as inputs, with the node degree $k$ sampled appropriately from the degree-weighted distribution $kp_k/c$. This process is then repeated until the statistics of the distribution $p(\omega)$ converge. At this point the total DOS $\rho(\lambda)$ can be evaluated as
\begin{equation}\label{total_DOS_pop_dyn}
\rho(\lambda) = \lim_{\varepsilon\to 0} \frac{1}{\pi} \mathrm{Re}\,\Big\langle \frac{\tau c}{\Omega_{k}(\lambda_{\varepsilon},\{\omega_l\},\tau)} \Big\rangle_{\{\omega_l\},\tau,k}
\end{equation}
where the angle brackets $\langle \ldots \rangle_{\{\omega_l\},\tau,k}$ indicate averaging over the degree distribution $p_k$, the lifetime distribution $\rho_\tau(\tau)$ and the $k$ cavity precision distributions\footnote{The full expression reads $\rho(\lambda) = \lim_{\varepsilon\to 0} \frac{1}{\pi} \mathrm{Re}\sum_k p_k\int\mathrm{d}\tau \rho_{\tau}(\tau) \,\prod_{l=1}^{k}\mathrm{d}\omega_l\,p(\omega_l)\,\tau c/\Omega_{k}(\lambda_{\varepsilon},\{\omega_l\},\tau)$} $\prod_{l=1}^{k}p(\omega_l)$. Comparing (\ref{total_DOS_pop_dyn}) with (\ref{total_DOS_from_conditional_DOS}) one reads off directly that the {\em local} DOS is given by an almost identical expression 
\begin{equation}\label{local_DOS_poulation_dynamics}
	\rho(\lambda|\tau) = \lim_{\varepsilon\to 0} \frac{1}{\pi} \mathrm{Re}\,\Big\langle \frac{\tau c}{\Omega_{k}(\lambda_{\varepsilon},\{\omega_l\},\tau)} \Big\rangle_{\{\omega_l\},k}
\end{equation}
that differs only in the fact that $\tau$ is fixed rather than averaged over.
The remaining average is, in practice, evaluated by sampling values of $k$ from the degree distribution, and sets 
$\{\omega_l\}$ of $k$ cavity precisions from the population $\mathcal{P}$ converged to equilibrium. 

\begin{figure}[htb]
	\centering
	\includegraphics{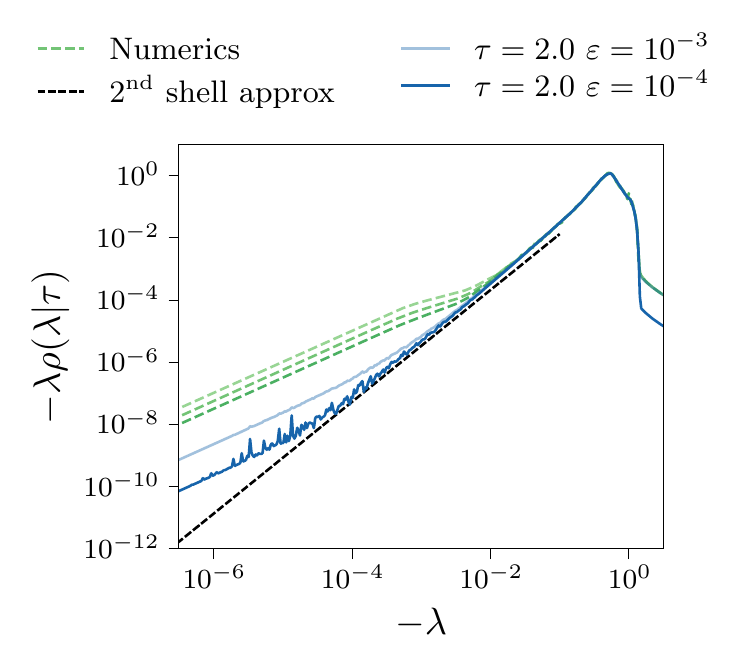}	\includegraphics{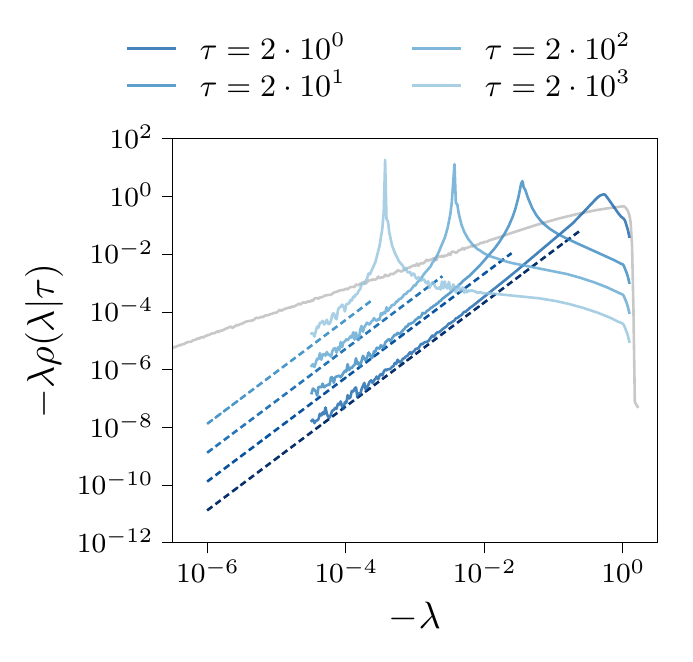}
	\caption{Left: local DOS for the random regular graph ensemble with $c = 5$, $T = 0.8$ and $\tau=2$. The blue lines show the results of the cavity method (population dynamics), while the green lines are obtained from direct diagonalization of systems with size $N=500,1000,2000$ (light to dark green), averaging across $10^4$ samples.  The black dashed line represents the second shell approximation given by Eq.~(\ref{second_order_approximation_small_lambda}). Right: local DOS for different values of $\tau$ (blue solid lines), and associated predictions of the second shell approximation (blue dashed lines, within the range $[10^{-5},1/(2\tau)]$). For these results we have used $\varepsilon$ ranging from $10^{-4}$ (for $\tau=2$) to $10^{-7}$  (for $\tau=2000$). The solid grey line in the background shows the total DOS.
	}\label{fig:local_DOS}
\end{figure}

Figure \ref{fig:local_DOS}-left shows the local DOS obtained by the above method on a log-log scale, for the random regular graph ensemble with $c=5$ and $T=0.8$, and for given $\tau=2$. The factor $-\lambda$ on the y-axis accounts for the transformation $\lambda\to \ln(-\lambda)$ on the x-axis, so this plot can be read as the distribution of the logarithmic relaxation rates, $\ln(-\lambda)$, with the correct normalization. The blue lines show the results obtained with the population dynamics algorithm described above, using two different values of $\varepsilon$. We note that a smaller value of $\varepsilon$ gives a better resolution on the negative lambda axis, as it should, and that straight $\varepsilon$-dependent tails appear in the small $|\lambda|$ region (below the corresponding values of $\varepsilon$), as well as for $|\lambda|>1$. These have no physical meaning: population dynamics sampling runs of finite length produce only a limited number of samples in this region, if any, and the resulting shape of the (local) DOS is strongly affected by the Lorentzians of width $\varepsilon$ that the method effectively uses to smooth the spectra. The green dashed lines were obtained from direct diagonalizations (labelled ``numerics" in the plots) of instances of $\mathbf{M}^{\textbf{s}}$ with network size $N=500, 1000, 2000$ (light to dark green). The agreement with the population dynamics results is excellent in the regions where finite size effects are absent. The latter do show up in the small $|\lambda|$ regime and are again a consequence of limited sample sizes, with finite matrices only rarely having eigenvalues in this region\footnote{The intuition here is that the eigenvalues determine the relaxation times of the system, and exploring a smaller network must take less time, on average. More precisely, the largest trapping time $\tau_{\text{max}}$ in a finite system of size $N$ scales as $\tau_{\text{max}}\sim N^{\beta}$ \cite{Margiotta2018}.}. 

In figure \ref{fig:local_DOS}-right we explore the dependence of the local DOS on $\tau$ as predicted by the population dynamics algorithm, for the same setting of random regular networks with $c=5$ at $T=0.8$. The grey solid line in the background is the total DOS given by (\ref{total_DOS_pop_dyn}). This plot provides two important insights about the local DOS: (i) most of its mass is peaked around a value of $-\lambda$ that scales as $1/\tau$, with this peak getting narrower as $\tau$ increases; (ii) in the small $|\lambda|$ regime well below the peak, the local DOS has a power law dependence on $|\lambda|$, with a $\tau$-independent exponent. 

\begin{figure}[htb]
	\centering
	\includegraphics{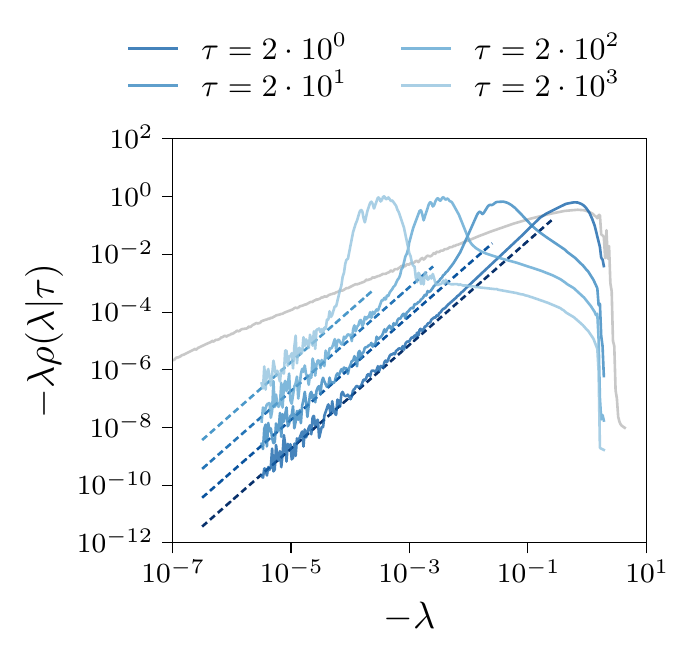}	\includegraphics{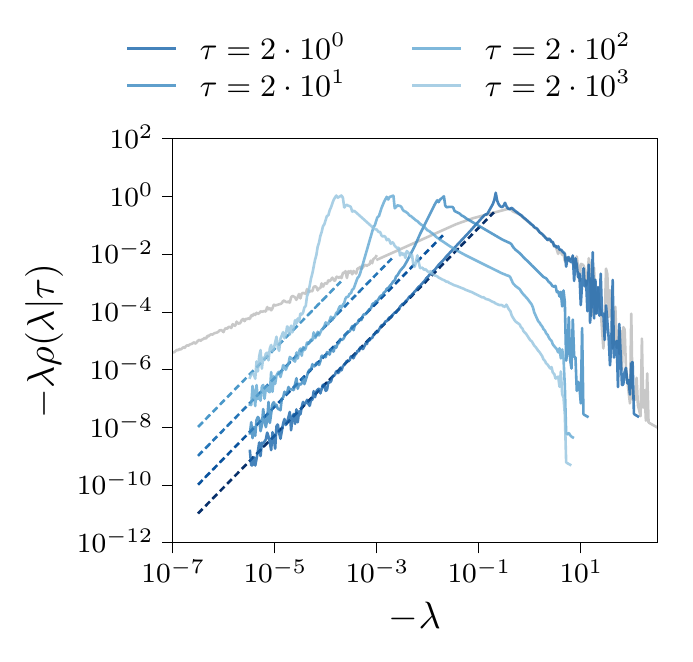}
	\caption{Left: local DOS for the Erd\"os-R\'enyi graph ensemble with $c=5$, $T=0.8$; the degree distribution is $p_k\propto c^{k}e^{-c}/k!$, with an upper bound $k_\mathrm{max}=100$ imposed for numerical efficiency. Blue solid lines: population dynamics results for different $\tau$, with $\varepsilon$ as in Fig.~\ref{fig:local_DOS}. Blue dashed lines: second shell approximation predictions. Solid grey line: total DOS. Right: analogous plot for the case of scale-free networks with degree distribution $p_k\propto k^{-\gamma}$, bounded between $k_\mathrm{min}=2$ and $k_\mathrm{max}=1000$; the exponent $\gamma=2.5$ and mean degree $\langle k \rangle = 4.54$ are chosen to match results for the configuration space topology of a system of Lennard-Jones particles \cite{Doye2002}.
	}\label{fig:local_DOS_ER_SF}
\end{figure}

At this stage it is useful to compare to the simpler MF ($c\to\infty$) and RW ($T\to\infty$) limits discussed in the introduction. We discuss their local DOS in Appendix \ref{appendix_localDOS_MF_RW} and find in both cases a delta peak at $|\lambda^*|\propto1/\tau$ for large $\tau$. (The proportionality constant is unity for the MF case, where the delta peak is the only contribution to the local DOS; in the RW case there is an additional piece to the spectrum for $|\lambda|$ of order unity.) The delta peak is the analogue of the smooth peaks visible in figure \ref{fig:local_DOS}-right. More importantly, the local DOS turns out to be {\em zero} for $|\lambda|$ below the peak. The power law behaviour in this regime that we see in figure \ref{fig:local_DOS}-right therefore has no analogue in either the MF or RW limits: it arises only as a {\em combined} effect of limited connectivity and trap depth disorder. Remarkably, the power law behaviour for small $|\lambda|$ is robust to changes in the network connectivity, as can be seen by comparing results for the  Erd\"os-R\'enyi and scale-free network ensembles in figure \ref{fig:local_DOS_ER_SF} to those for random regular networks in \ref{fig:local_DOS}-right. One observes that while the shape of $\rho(\lambda|\tau)$ around $\lambda = -1/\tau$ does depend on the type of network, the exponent of the power law tail for low $|\lambda|$ does not.

In the next section we present an approximation scheme that can explain the power law behaviour observed in the local DOS for $|\lambda|\ll1/\tau$, and clarifies that this result applies to any sparse network specified by some degree distribution $p_k$ with finite mean. In fact, the argument that we use is insensitive to correlations among node degrees, and should therefore remain valid for networks generated e.g.\ by preferential attachment \cite{Albert2002}.
%
%
\subsection{Approximation scheme}\label{subsection_approximation_scheme}
While the equations resulting from the cavity method do not permit closed-form solutions, we can use them to construct an approximation scheme that provides useful insights. This is done in the spirit of the single defect approximation \cite{Biroli1999, Semerjian2002}: the main idea is to take into account the disorder of a certain region of interest only, e.g.\ a single node or a given neighbourhood in the case of a network, and assume the rest of the system to be homogeneous. In our model, a first order (or ``first shell'') approximation of this kind corresponds to assuming $T=\infty$ and a $c$-regular connectivity on the cavity network, and taking only the lifetime $\tau$ and connectivity $k$ of the central node into account. With these assumptions the cavity network becomes disorder-free in the thermodynamic limit, which implies that the distribution $\rho(\omega)$ of cavity precisions becomes a delta function centred on the value of $\bar{\omega}$ that solves $\bar{\omega}=\Omega_{c-1}(\lambda_{\varepsilon},\{\bar{\omega}\},1)$. We refer to $\bar{\omega}$ as the infinite-$T$ (and $c$-regular) solution. The first order approximation is then implemented as one cavity step -- involving the local $\tau$ and $k$ values -- performed starting from the infinite-$T$ solution. Following this idea, the second order (or second shell) approximation consists of two cavity steps from the infinite-$T$ solution and so on. In what follows we will focus on the random regular graph ensemble for simplicity, and refer to the appendices for the demonstration of the wider applicability of the approximation.

The first order approximation was previously found to give an accurate description of the total DOS in the small $|\lambda|$ regime, where $\rho(\lambda)\sim (-\lambda)^{T-1}$ as in mean field \cite{Bovier2005, Margiotta2018}. For the local DOS, on the other hand, the first order approximation is not able to provide a match to the power law behaviour shown in figure \ref{fig:local_DOS}. In fact, this approximation coincides with the random walk limit discussed in appendix \ref{appendix_localDOS_MF_RW}, which yields a local DOS characterised by a single delta peak in the small $|\lambda|$ regime. This is due to the complete lack of randomness in the approximation: the attributes of the central node are fully specified (by $\tau$ and the fixed degree $k=c$) as are those of the cavity network by the infinite-$T$ solution. The second order approximation is then required if we want to include some of the original heterogeneity of the system: in the thermodynamic limit, different nodes with the same value of $\tau$ have different neighbourhoods, and averaging across these neighbourhoods gives the approximated local DOS. This can be expressed mathematically as 
\begin{equation}\label{second_order_approximation}
\rho^{\mathrm{2A}}(\lambda|\tau)=\lim_{\varepsilon\to 0}\frac{1}{\pi}\text{Re}\Big\langle \frac{\tau c}{\Omega_k(\{\Omega_{c-1}(\{\bar{\omega}\},\tau_l)\},\tau)} \Big\rangle_{\{\tau_l\},k}
\end{equation}
where the superscript 2A stands for ``second order approximation". In appendix \ref{appendix_second_shell_approximation} we show that for $|\lambda|\ll 1/\tau$ Eq.~(\ref{second_order_approximation}) can be written as
\begin{equation}\label{second_order_approximation_delta}
\rho^{\mathrm{2A}}(\lambda|\tau)\simeq c\tau \Big\langle \delta(k-\sum_l^k y_l) \Big\rangle_{\{y_l\},k}
\end{equation}
with $y_l = 1/(\lambda c [\tau_l+(c-2)^{-1}]+c-1)$. The distribution $\rho_y(y_l)$ of $y_l$ is strongly peaked within a region of order $|\lambda|^T$ around $1/(c-1)$, and it drops by a factor $|\lambda|^T$ if $y_l$ is away from this value. The average on the right hand side of (\ref{second_order_approximation_delta}) is then dominated by cases where all the $y_l$ are close to $1/(c-1)$ except for one, say $y_1$. The latter then has to equal $k-(k-1)/(c-1)$, which happens when the corresponding $\tau_1$ is of order $1/|\lambda|$. Substituting the expression for $\rho_y(y_1=k-(k-1)/(c-1))$ and multiplying by a factor $k$ (as any of the $\tau_l$ could be the large one) this leads to the following result:
\begin{equation}\label{second_order_approximation_small_lambda}
\rho^{\mathrm{2A}}(\lambda|\tau)\approx \tau C(c,T)|\lambda|^{T}\quad \text{for}\quad |\lambda|\ll 1/\tau
\end{equation}
where the prefactor $C(c,T)$ depends on the degree distribution of the network and vanishes in the limit $c\to\infty$, as required to recover the mean field case where the local DOS vanishes for $|\lambda| \ll 1/\tau$. For random regular networks we find explicitly $C(c,T)=Tc(c-1)^{T-1}(c-2)^{-(T+1)}$. Equation (\ref{second_order_approximation_small_lambda}) then also implies $\rho^{\mathrm{2A}}(\lambda|\tau)\to 0$ when $T\to\infty$, at least for $|\lambda|<(c-2)/(c-1)$, and so -- like the first order approximation -- it is consistent with the random walk limit. The predicted small-$|\lambda|$ power law of the local DOS, $\rho(\lambda|\tau)\sim |\lambda|^T$, matches the full population dynamics results well (see the black dashed line named ``$2^{\text{nd}}$ shell approximation" in figure \ref{fig:local_DOS}-left, and the blue dashed lines in figures \ref{fig:local_DOS}-right and \ref{fig:local_DOS_ER_SF}). Interestingly, the argument that leads to Eq.~(\ref{second_order_approximation_small_lambda})  implies that the observed power law arises because of deep minima surrounding the node of interest: a significant contribution to the local DOS at any given $\lambda$ comes from those traps that have at least one neighbouring minimum with expected lifetime $\gtrsim 1/|\lambda|$. As we will see, this feature of the local DOS determines the observed long time behaviour of several quantities of interest, and in particular it has important implications for the return probability discussed in the next section.


\section{Return probability}\label{section_return_probability}
We now turn our attention to the time domain. We start by using the results for the local DOS to compute the average probability of return to an initial trap, as a function of its lifetime $\tau$.

Let us call $P_i(t)$ the return probability to an arbitrary initial trap $i$. Physically, this gives the probability of being in trap $i$ at time $t$, given that we have started out in the same trap. Mathematically it is given by equation (\ref{p_t}) with the initial condition $p_j(0)=\delta_{ij}$, yielding
\begin{equation}\label{return_probability}
P_{i}(t) =  \sum_{\alpha} e^{\lambda_{\alpha}t}  u_{\alpha, i}w_{\alpha, i} = \sum_{\alpha} e^{\lambda_{\alpha}t}  v_{\alpha,i}^2  
\end{equation}
In the second equality we have used the relations $u_{\alpha, i}=p_{{\rm eq},i}^{1/2}\, v_{\alpha, i}$ and $w_{\alpha, i}=p_{{\rm eq},i}^{-1/2} v_{\alpha, i}$ to transform to an expression in terms of the eigenvectors of the symmetrized master operator; note that the symmetrization factors cancel. Decomposing the sum according to the eigenvalues $\lambda$ gives
\begin{equation}
P_{i}(t)= \int \mathrm{d}\lambda \, \sum_{\alpha}  \delta(\lambda-\lambda_{\alpha})  v_{\alpha,i}^2  e^{\lambda t} = \int \mathrm{d}\lambda \, \rho(\lambda|i) e^{\lambda t}
\end{equation}
where $\rho(\lambda|i)$ is the single node local DOS defined in (\ref{local_DOS_single_site}). We now define an average return probability over all traps with lifetime $\tau_i=\tau$:
\begin{equation}\label{return_probability_from_localDOS}
P_{\tau}(t)= \frac{1}{N_\tau} \sum_{i=1}^{N} \delta(\tau-\tau_i)P_i(t)= \int \mathrm{d}\lambda \, \rho(\lambda|\tau) e^{\lambda t}
\end{equation}
As the second equality shows, the return probability $P_{\tau}(t)$ can be obtained directly from the local DOS $\rho(\lambda|\tau)$, which in turn we can predict using population dynamics as explained above. We can also use  (\ref{return_probability_from_localDOS}) in reverse to deduce that the average of $\lambda$ over the local DOS $\rho(\lambda|\tau)$ is given by $-1/\tau$, in accordance with our findings about the scaling of the peaks observed in figure \ref{fig:local_DOS}. This general result for the average of $\lambda$ can be seen by noting that $P_i(t)=(\exp \mathbf{M} t)_{ii}=1+M_{ii}t+O(t^2)$. From the definition (\ref{master_operator}) of the master operator this equals $1-t/\tau_i+O(t^2)$; comparing with the expansion of the r.h.s.\ of (\ref{return_probability_from_localDOS}) to linear order in $t$ then gives the result.

\begin{figure}[htb]
	\centering
	\includegraphics{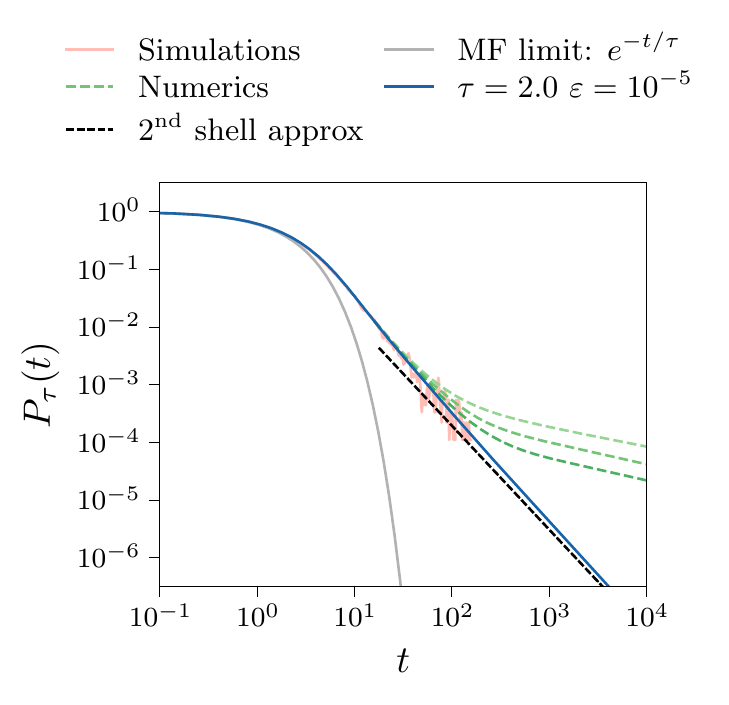}
	\caption{Average return probability for dynamics on random regular networks with $c=5$ and $T=0.8$, for initial traps with expected lifetime $\tau=2$. Blue line: result from numerical evaluation of Eq.~(\ref{return_probability_from_localDOS}) using the population dynamics data for the local DOS.  
		Black dashed line: second shell approximation for $t\gg\tau$ (see Eq.~(\ref{second_order_approximation_return_probability})). Green dashed lines: results from direct diagonalizations of the master operator for system sizes $N=500,1000,2000$ (light to dark green), averaging across $10^4$ samples. Red line: result from stochastic simulations using the Gillespie algorithm. Grey line: staying probability $S_{\tau}(t)=e^{-t/\tau}$.
	}\label{fig:p_vs_t}
\end{figure}

Figure \ref{fig:p_vs_t} shows the return probability $P_{\tau}(t)$ for the random regular graph ensemble with $c=5$, $T=0.8$ and $\tau=2$ obtained using different approaches: the blue line is computed from the population dynamics result discussed in the previous section, the green dashed lines are obtained using data from direct diagonalizations of samples of $\mathbf{M}^{\mathrm{s}}$-matrices of different sizes (light to dark green), and the red line shows the result of stochastic simulations with the Gillespie algorithm (see appendix \ref{appendix_simulations}). From short times up to $t\sim 10^1$ the agreement between these three approaches is clearly very good. At larger $t$, finite size effects induce upward curvature in the direct diagonalization curves as the estimate of $P_\tau(t)$ approaches the Boltzmann equilibrium distribution in a finite system. The simulations, which are performed on effectively infinite networks and so do not suffer from analogous errors, start to exhibit statistical sampling fluctuations in the same range of times. (With the $10^5$ runs of the simulated dynamics that we use, a reasonable estimate of the return probability can only be obtained for $P_{\tau}(t)\gtrsim 10^{-4}$; we therefore do not show data beyond this point in Fig.~\ref{fig:p_vs_t}.)

Looking now in more detail at the results in Figure \ref{fig:p_vs_t}, one observes that for times $t$ up to the order of the trap lifetime $\tau$, $P_{\tau}(t)$ is dominated by the ``staying probability'' of having never left the initial trap, which we will denote by $S_{\tau}(t)$. This quantity coincides with $P_{\tau}(t)$ for the fully-connected graph, which is the original Bouchaud model: once the initial trap has been left, the probability of coming back is $O(1/N)$ and vanishes in the thermodynamic limit. In our continuous time framework the staying probability is simply given by (see grey solid line in figure \ref{fig:p_vs_t})
\begin{equation}\label{staying_probability}
S_{\tau}(t)=e^{-t/\tau}
\end{equation}
The power law tail of the return probability observed beyond this, in the long time regime ($t>\tau$), is therefore clearly a network effect in the dynamics: the finite connectivity allows for returns to the initial trap even in the infinite system size limit, and as we will see this generates a power law tail provided that $T$ is finite so that the disorder in the trap depths matters. The power law predicted by the cavity theory is consistent with our simple estimate (\ref{second_order_approximation_small_lambda}) of the local DOS for $|\lambda|\ll 1/\tau$, which when inserted into the integral in (\ref{return_probability_from_localDOS}) yields
\begin{equation}\label{second_order_approximation_return_probability}
	P_{\tau}(t)\approx \tau \tilde{C}(c, T) \, t^{-(T+1)} \quad \text{for}\quad t\gg \tau
\end{equation}
The prefactor here is $\tilde{C}(c, T)= C(c, T)\,\Gamma(T+1)$, with $\Gamma(z)$ the Gamma function. This approximation agrees with the cavity theory exactly regarding the power law exponent; even the prefactor is quantitatively close (compare the blue solid line and the black dashed line in figure \ref{fig:p_vs_t}). 

In the previous section we explained that the derivation of (\ref{second_order_approximation_small_lambda}) implies that the small $|\lambda|$ power law tail of the local DOS originates from deep minima (with lifetimes $\gtrsim 1/|\lambda|$) surrounding the initial trap. The implication for the return probability is the following: the most likely manner in which the system can return to the initial state at $t\gg\tau$ is for it to become trapped in a neighbouring node with lifetime $\sim t$. Other possibilities of course exist, e.g.\ the system could come back from a trap in the second neighbour shell, but these only make a sub-dominant contribution to $P_{\tau}(t)$. This can be seen indirectly from the fact that our approximation almost overlaps with the numerically exact population dynamics curve in figure \ref{fig:p_vs_t}.

\begin{figure}[htb]
	\centering
	\includegraphics{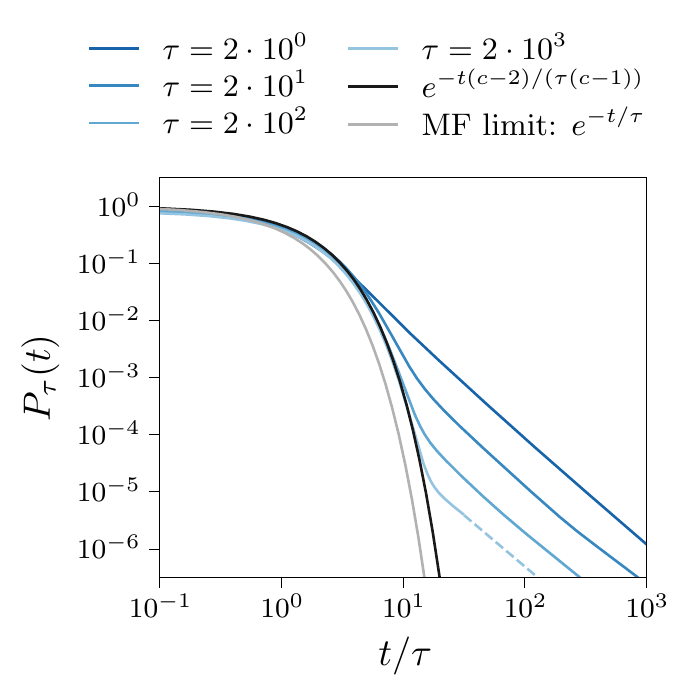}	\includegraphics{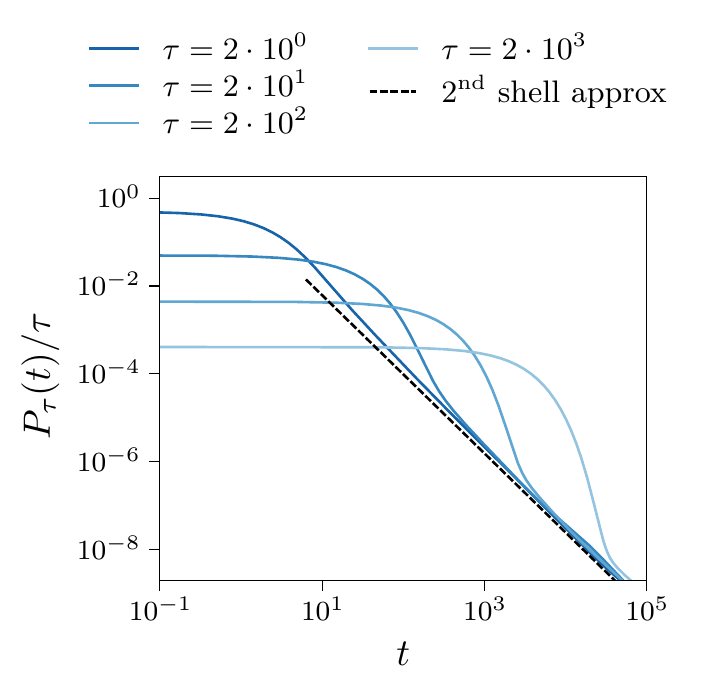}
	\caption{Left: average return probability $P_{\tau}(t)$ against $t/\tau$ for dynamics on random regular networks with $c=5$, at temperature $T=0.8$. Blue lines: results for different $\tau$, obtained from population dynamics data for the local DOS. The collapse for $t/\tau\leq 1$ follows the staying probability (grey line) except for a $c$-dependent factor in the decay rate (see main text). Right: analogous plot of $P_{\tau}(t)/\tau$ against $t$, giving a collapse in the power law tail for $t\gg \tau$, as predicted by the second shell approximation (see Eq.~(\ref{second_order_approximation_return_probability}))  .}\label{fig:p_vs_t_all_tau}
\end{figure}

Summarizing, $P_{\tau}(t)$ exhibits an initial exponential decay typical of the mean field limit for $t\lesssim \tau$, followed by a power law behaviour with a $T$-dependent exponent for $t\gtrsim \tau$; the latter arises from deep minima surrounding the departing node. Note that the crossover point from the exponential decay to the power law regime occurs at a value of $P_\tau(t)$ that decreases as $\tau$ increases, which is directly related to the fact that the power law tail does not just depend on the scaled time $t/\tau$ (see (\ref{second_order_approximation_return_probability})). Both features can be seen in figure \ref{fig:p_vs_t_all_tau}, where we compare evaluations of $P_{\tau}(t)$ for different values of $\tau$: in the left plot the x-axis is scaled by $\tau$, which delivers a collapse of the exponential decay at short times, while in the right plot we have similarly scaled the y-axis to show that the prefactor of the tail is indeed proportional to $\tau$ as the approximation (\ref{second_order_approximation_return_probability}) predicts. Note that the time constant of the initial exponential decay is somewhat larger than the mean field value $\tau$. In fact for $\tau\gg1$ this can be approximated by $\tau (c-2)/(c-1)$, as we will justify in the next section.

We note finally that the conditioning on the trap lifetime in the return probability $P_{\tau}(t)$ is essential in order to isolate the effects of finite, non-mean field network connectivity among traps. If one instead considers the probability of return to a {\em randomly} chosen initial trap $P(t)=\int \mathrm{d}\tau\,P_{\tau}(t)\rho_{\tau}(\tau)$, one finds that this is dominated by the initial exponential decay of $P_\tau(t)$, which is exactly the mean field result. The long time scaling $P(t)\sim t^{-T}$ that one deduces has a mean field form even for finite connectivity $c$. The decay exponent is consistent with the small $\lambda$-scaling of the total DOS \cite{Bovier2005, Margiotta2018}, which again is independent of $c$. This analysis tells us that the trap model on networks exhibits a long time mean field dynamics \emph{on average} only, with network effects coming to the fore when considering more detailed phenomena like returns to specific initial traps as considered here.


\section{Excursion times}\label{section_excursion_times}

In the previous section we discussed the dynamical properties of the system in terms of the return probability to some initial trap. We saw that the local disorder, i.e.\ the depth of the departing trap, determines the shape of $P_{\tau}(t)$ at short times. In contrast, the long time behaviour is always a power law in $t$, with the dependence on $\tau$ entering only via the prefactor. We suggested that this is because when $t$ is much larger than the trap lifetime $\tau$, the probability of finding the system in the original trap is dominated by the lifetimes of the {\em neighbouring} minima. In this section we consolidate this idea by excluding the time spent in the initial trap from the analysis and looking at the distribution of excursion times, i.e.\ the time spent by the system away from the initial trap between two visits there. To this end we decompose the return probability as
\begin{equation}\label{return_probability_decomposed}
P_{\tau}(t) = \sum_{n=0}^{\infty} P_{\tau}^{(n)}(t)
\end{equation}
where $P_{\tau}^{(n)}(t)$ indicates the probability of finding the system in the original trap at time $t$, assuming that it has returned there $n$ times in total. Note that $P_{\tau}^{(0)}(t)=S_{\tau}(t)$, while $P_{\tau}^{(1)}(t)$ can be written as
\begin{equation}\label{return_probability_1_return}
P_{\tau}^{(1)}(t) = \int_{t_1}^{t}\mathrm{d}t_2\int_{0}^{t}\mathrm{d}t_1 \, L_{\tau}(t_1) F(t_2-t_1) S_{\tau}(t-t_2)
\end{equation}
Here the integrand is the probability that the system leaves the initial trap at time $t_1$ (we write this as $L_{\tau}(t_1)\,\mathrm{d}t_1$), stays away until time $t_2$ ($F(t_2-t_1)\,\mathrm{d}t_2$), then returns to the origin and remains there until $t$ ($S_{\tau}(t-t_2)$). Reading the expression in this way implies that $F(t)$ is the distribution of excursion times, the quantity that we want to evaluate.  Equation (\ref{return_probability_1_return}) is the convolution between $L_{\tau}$, $ F$ and $S_{\tau}$, i.e. $P_{\tau}^{(1)} = L_{\tau} \ast  F \ast S_{\tau}$, which can be generalised to $n$ returns as
\begin{equation}\label{return_probability_n_conv}
P_{\tau}^{(n)} = (L_{\tau} \ast  F )^{(n)}\ast S_{\tau}
\end{equation}
with $(L_{\tau} \ast  F )^{(n)}$ the $n$-fold convolution between $L_{\tau}$ and $ F$. Substituting (\ref{return_probability_n_conv}) into (\ref{return_probability_decomposed}) and taking the Laplace transform leads to
\begin{equation}\label{return_probability_laplace_transform}
\hat{P}_{\tau}(s) = \frac{\hat{S}_{\tau}(s) }{1-\hat{L}_{\tau} (s) \hat{F} (s)}
\end{equation}
which can be written more explicitly using that, from (\ref{staying_probability}),
\begin{equation}\label{staying_probability_laplace_transform}
	\hat{S}_{\tau}(s)=\tau/(1+s\tau)
\end{equation}
Also one has $L_{\tau}(t)=-S'_{\tau}(t)$ or in Laplace space $\hat{L}_{\tau}(s) = 1-s \hat{S}_{\tau}(s)=1/(1+s\tau)$. Substituting into (\ref{return_probability_laplace_transform}) 
yields
\begin{equation}\label{return_probability_laplace_space}
\hat{P}_\tau(s) = \frac{\tau}{1+s\tau-\hat{F}(s)}
\end{equation}
This equation can be inverted to obtain an expression for the excursion time distribution in terms of the return probability,
\begin{equation}\label{excursion_time_probability_laplace_transform}
\hat{F} (s) = s \tau + 1 - \tau / \hat{P}_{\tau}(s)
\end{equation}

We consider first the limit $s\to 0$. The value $\hat{F} (0)=\int_{0}^{\infty} \mathrm{d}t\,F(t)$ gives the probability that an excursion lasts \emph{any finite} amount of time, i.e.\ the probability that the system will sooner or later go back to the initial node rather than escape to infinity. On the infinite $c$-regular tree, i.e.\ for a random $c$-regular graph in the limit $N\to\infty$, this fixes\footnote{A simple argument to see this is the following. The probability of ever returning -- call this $P_0$ -- cannot depend on the lifetimes of the traps in the configuration space: if the system escapes to infinity it does not matter how long this will take, and so $P_0$ is independent of $T$ as long as $T>0$. Moreover we have $P_0=P_{01}$, where $P_{nm}$ is the probability to ever land on a node in the $n^{\mathrm{th}}$ neighbour shell of the initial trap, starting from the $m^{\mathrm{th}}$ shell (here the $n=0$ ``shell'' is the initial trap). It is immediate to see that $P_{01} = 1/c + P_{02}(c-1)/c $ and also $P_{02}=P_{01}P_{12}$, which by symmetry becomes $P_{02}=P_{01}^2$. The resulting second order equation gives the physical solution $P_{01}=1/(c-1)$, which correctly becomes $P_{01}=1$ for a chain ($c=2$) and $P_{01}=0$ in the MF limit ($c\to\infty$). In general, the probability to reach a node that is $l$ steps away decreases exponentially, $P_{0l}=(c-1)^{-l}$, which suggests a possible explanation for why the $2^{\mathrm{nd}}$ shell approximation works so well.} $\hat{F}(0) = 1/(c-1)$, and so we obtain $\hat{P}_\tau(0)=\tau (c-2)/(c-1)$ from (\ref{excursion_time_probability_laplace_transform}), while $\hat{S}_{\tau}(0)=\tau$. In the MF limit we have $\hat{P}_{\tau} (s) = \hat{S}_{\tau} (s)$, implying $\hat{F}_{\tau} (s) =0$ via (\ref{excursion_time_probability_laplace_transform}), which is consistent with the $c\to\infty$ limit of $\hat{F}(0)=1/(c-1)$.

Next we analyse what (\ref{excursion_time_probability_laplace_transform}) says about the long time behaviour of $F(t)$, by considering $\hat{F}(s)$ for small $s$. From equation (\ref{second_order_approximation_return_probability}) one obtains the following approximation for the leading singular small $s$-behaviour of the return probability in Laplace space\footnote{
One has generally $\hat{P}_{\tau}(0)-\hat{P}_\tau(s) = \int_{0}^{\infty}\mathrm{d}t\,P_{\tau}(t)(1-e^{-st})$. The integrand becomes dominated by large $t$ for small $s$; substituting the tail estimate (\ref{second_order_approximation_return_probability}) and integrating by parts then gives  (\ref{difference_return_prob_LT}).}
\begin{equation}\label{difference_return_prob_LT}
\hat{P}_{\tau}(s) - \hat{P}_{\tau}(0) \simeq \tau \Gamma(-T)\tilde{C}(c, T) \, s^{T} \quad \text{for}\quad s\ll 1/\tau
\end{equation}
Substituting this expression for $\hat{P}_{\tau}(s)$ into (\ref{excursion_time_probability_laplace_transform}) and expanding again for small $s$ one sees that $\hat{F}(s)$ contains the same singular term:
\begin{equation}\label{excursion_time_probability_laplace_transform_expansion}
\hat{F} (s) - \hat{F} (0) \simeq
\tau^2\,\frac{\Gamma(-T)\tilde{C}(c, T)}{\hat{P}^2_{\tau}(0)} s^T \quad \text{for}\quad s\ll 1/\tau
\end{equation}
For the long time behaviour of $F(t)$ this implies the same power law that we found in the return probability:
\begin{equation}\label{excursion_time_probability_long_time}
F(t) \simeq \Big(\frac{c-1}{c-2}\Big)^2 \tilde{C}(c, T) \, t^{-(T+1)}
\end{equation}
where we have used $\hat{P}_{\tau}(0) = \tau(c-2)/(c-1)$. Note that the predicted behaviour of the excursion time distribution only depends on the average connectivity and temperature, while the departing lifetime $\tau$ disappears from (\ref{excursion_time_probability_long_time}) as it did in our earlier result $\hat{F}(0)=1/(c-1)$. This is as expected: in the Bouchaud trap model, the escape time $\tau$ only contains information on the {\em local} disorder (trap depth) at the initial minimum, so once this trap has been left, the behaviour during the following excursion is independent of $\tau$.

So far our analytical reasoning was based on an approximation for the local DOS, which we converted into a return probability $P_\tau(t)$ and finally into the excursion distribution $F(t)$. Qualitatively, we can also alternatively argue directly from $F(t)$, by constructing a simple lower bound. The probability of an excursion taking longer than $t$, $\int_t^\infty \mathrm{d}t' F(t')$, is at least as large as the probability of not having left the first trap encountered during the excursion. As the depth of this trap is random, the latter probability is $\int \mathrm{d}\tau'\, \rho_\tau(\tau') e^{-t/\tau'}$. This lower bound is just the {\em mean field} return probability $P(t)\sim t^{-T}$ discussed at the end of the previous section. Taking a derivative w.r.t.\ $t$ gives the estimate that $F(t)$ should decay as $t^{-(T+1)}$, exactly as we had found in  (\ref{excursion_time_probability_long_time}). This then implies the analogous power law (\ref{second_order_approximation_return_probability}) in the return probability $P_{\tau}(t)$, and in turn via (\ref{return_probability_from_localDOS}) the small $|\lambda|$ power law tail (\ref{second_order_approximation_small_lambda}) we observed in the local DOS. Note that the above bound for the cumulative excursion time distribution again supports our intuitive ``deep minimum in the first shell'' picture: the average of $e^{-t/\tau'}$ is dominated by traps with lifetimes $\tau' \gtrsim t$, i.e.\ by the deepest minima surrounding the initial trap.

Before showing numerical results we comment briefly on the short time behaviour of $F(t)$. This is determined by the average escape rate $\int \mathrm{d}\tau'\, \rho_\tau(\tau')/\tau'$ of the first neighbour traps, which have random depths, multiplied by the probability $1/c$ (again for the random regular graph ensemble) of making the first jump from such a neighbour back to the initial trap. Overall this yields $F(0)=T/(c\,(T+1))$. This constant translates into a $1/s$ power law in Laplace space for $s\gg1$; such a power law also appears in $\hat{P}_{\tau}(s)$ due to $P_{\tau}(0)=1$.

\begin{figure}
	\centering
	\includegraphics{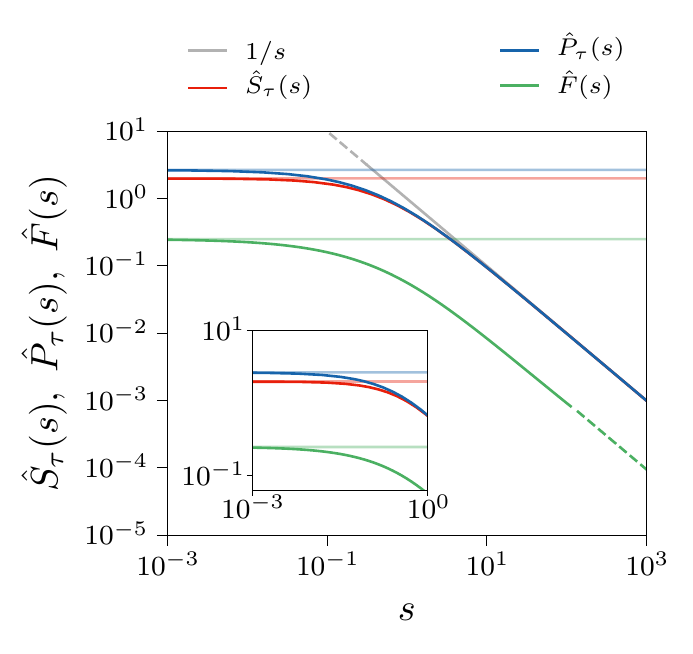}\includegraphics{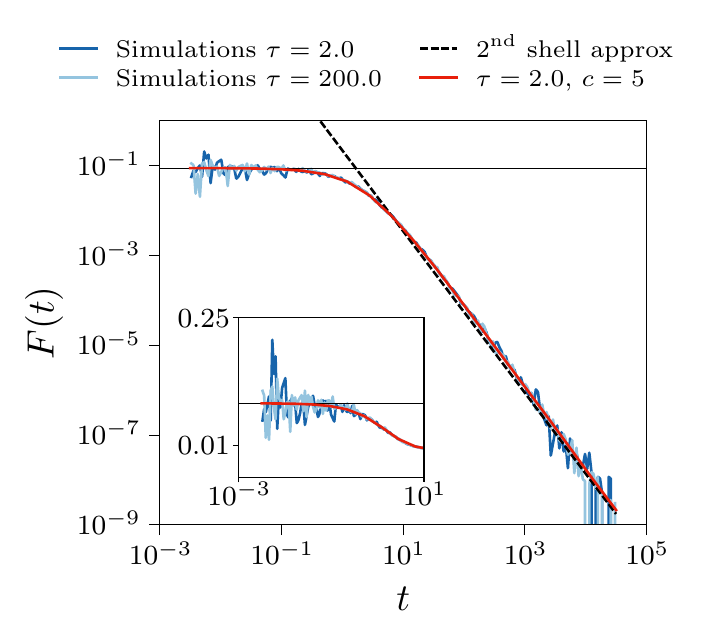}
	\caption{Left: Laplace transforms of the return probability (blue line), staying probability (red line) and excursion time distribution (green line) for the random regular graph ensemble with $c=5$, $T=0.8$ and departing traps with $\tau=2$. The dashes continuing the green line for $s>10^2$ represent the $1/s$ behaviour of $\hat{F}(s)$ at large $s$ (the evaluation via (\ref{excursion_time_probability_laplace_transform}) drops below numerical precision here as $\hat{P}_{\tau}(s)$ exhibits the same behaviour - see blue and grey lines). The horizontal lines correspond to $\hat{P}_{\tau}(0)$ (blue), $\hat{S}_{\tau}(0)$ (red) and $\hat{F}(0)$ (green). Inset: zoom in on the small $s$ range. Right: excursion time distribution obtained by inverse Laplace transform of $\hat{F}(s)$ from the left plot (red line). The black dashed line is the theoretical estimate given by Eq.~(\ref{excursion_time_probability_long_time}). The plot also shows $F(t)$ estimated from simulations on the infinite tree which are robust to changes in $\tau$.}\label{fig:P_F_S}
\end{figure}

Our numerical results for the excursion time distribution and the related quantities are displayed in figure \ref{fig:P_F_S}: on the left we have the staying probability (red line), the return probability (blue line) and the excursion time distribution (green line) in Laplace space for the random regular graph ensemble with $c=5$, $T=0.8$ and considering departing traps with lifetime $\tau=2$. $\hat{P}_{\tau}(s)$ and $\hat{F}_{\tau}(s)$ are computed using the population dynamics results for the local DOS, while $\hat{S}_{\tau}(s)$ is given explicitly by Eq.~(\ref{staying_probability_laplace_transform}). Note that all these quantities decrease as $1/s$ for large $s$ as expected. The inset shows the small $s$ behaviour where the horizontal lines correspond to the values $\hat{P}_{\tau}(0)$, $\hat{F}(0)$ and $\hat{S}_{\tau}(0)$ derived above. The plot on the right has the excursion time distribution $F(t)$ evaluated as the inverse Laplace transform of the data in the left plot, together with the results from direct simulations on the infinite $c$-regular tree (blue lines); we show simulations for multiple $\tau$, which produce identical results as expected. These numerical results match the approximation (\ref{excursion_time_probability_long_time}) well, see the black dashed line labelled $2^{\mathrm{nd}}$ order approximation. The inset clarifies the behaviour of $F(t)$ for $t\to 0$, with the horizontal line indicating the asymptote $F(0)$.

We stress that our results on the long time power law behaviour of the return probability and excursion time distribution are robust to changes in network topology, as long as this exhibits a locally tree-like structure. Our deep minima argument continues to apply then, with returns from the first neighbour shell surrounding the initial node giving the dominant contribution.

We return finally to the return probability $P_\tau(t)$ as shown in Fig~\ref{fig:p_vs_t_all_tau}. As discussed, this quantity initially decays exponentially in $t/\tau$, and the range where this decay is seen expands without bound as $\tau\to\infty$. The decay constant is somewhat slower than the staying probability $S_\tau(t)$ would suggest, however. To understand this, one can focus on the $t/\tau$ scaling of $P_\tau(t)$ by considering in Laplace space $\tau^{-1}\hat{P}_\tau(s)$ for $s=\sigma/\tau$. From (\ref{return_probability_laplace_space}) this is just $\tau^{-1} \hat{P}_\tau(\sigma/\tau) = [1+\sigma-\hat{F}(\sigma/\tau)]^{-1}$. In the large $\tau$-limit that we are interested in, $\hat{F}(\sigma/\tau)\to \hat{F}(0)=1/(c-1)$ so that $\tau^{-1} \hat{P}_\tau(\sigma/\tau) \to [(c-2)/(c-1)+\sigma]^{-1}$. This implies in the time domain in the same limit that
\begin{equation}
P_\tau(t) = e^{-(t/\tau)(c-2)/(c-1)}
\label{return_prob_large_tau}
\end{equation}
 which is consistent with the numerical results shown in Fig.~\ref{fig:p_vs_t_all_tau}-left. The exponential decay resembles that of the staying probability $S_\tau(t)=\exp(-t/\tau)$, but is somewhat slower except in the mean field limit $c\to\infty$. The slowing down arises from the fact that the system can return an arbitrary number of times $n$ to the initial trap, and the sum of all the contributions from $n=1, 2, \ldots$ returns just conspires to produce an exponential return probability decay with a smaller decay rate. An intuitive physical explanation can be formulated as follows: when $\tau\gg1$, excursions take negligible time compared to $\tau$. Since $\hat{F}(0)$ is the return probability, the probability of escaping in any attempt is only $1-\hat{F}(0)$. Therefore the rate of escape is $(1-\hat{F}(0))/\tau=(c-2)/(c-1)\tau^{-1}$.


\section{Conclusion}\label{section_conclusion}

In this paper we have studied a model for the evolution of glasses in configuration space: the dynamics takes place on a random network whose nodes represent the energy minima (or traps) of the system. These are characterised by the number of nearest neighbours $k$ and an average lifetime $\tau$. The latter is a quenched random variable, power law distributed, whose average diverges below the glass transition temperature $T=1$. Our focus was on the spectrum of eigenvalues $\lambda$ of the master operator governing the dynamics, and in particular on the local density of states $\rho(\lambda|\tau)$, i.e.\ the contribution to the spectrum of relaxation rates from all traps with a fixed lifetime $\tau$.

We employed the cavity method to exploit the tree-like structure of infinite random networks, and computed numerically the local DOS using a population dynamics algorithm. The cavity construction also allowed us to perform a simple analytical approximation that provides a very good match to the exact numerical results: the local DOS shows a small-$|\lambda|$ power law tail governing the long time dynamics, specifically $\rho(\lambda|\tau)\sim\tau|\lambda|^T$ for $|\lambda|\ll1/\tau$. This result is robust to changes in the network topology as long as a locally tree-like structure is retained; in this class of networks we considered here random $c$-regular, scale-free and Erd\"os-R\'enyi graphs. The power law tail of the local DOS is associated, in the time domain, with the distribution $F(t)$ of excursion times $t$ away from some initial trap. We found $F(t)\sim t^{-(T+1)}$. This can be seen as responsible for the long time behaviour of the probability to return to traps of depth $\tau$, which is $P_{\tau}(t)\sim\tau t^{-(T+1)}$ for $t\gg\tau$.

We showed that the above dynamical properties arise as a combined effect of a sparse (loop-free) configuration space connectivity and the quenched trap depth disorder: when these features are considered separately the local DOS becomes $\delta$-shaped for small $|\lambda|$, implying an exponential decay of $F(t)$ and $P_{\tau}(t)$.
In more detail, our analysis indicates that the most likely way for the system to return to the departing trap at some time $t\gtrsim\tau$ is to spend most of the interim stuck in a neighbouring trap with lifetime of order $t$ or larger. Returns from more distant minima are possible but become exponentially less probable with distance. For $t\lesssim\tau$, instead, $P_{\tau}(t)$ is dominated by the probability of having never left the original trap, $S_{\tau}(t)=e^{-t/\tau}$; in the mean field limit of infinite connectivity, $S_\tau(t)$ is in fact the only contribution to $P_\tau(t)$. Finally, the exponential shoulder of $P_{\tau}(t)$ dominates the {\em average} return probability $P(t)=\int\mathrm{d}\tau \rho_{\tau}(\tau)P_{\tau}(t)$, leading to the mean field scaling $P(t)\sim t^{-T}$, in accordance with the small $|\lambda|$ tail observed in the total DOS \cite{Margiotta2018}. This tells us that the long time dynamics of the trap model on random networks is of mean field kind \emph{on average} only, while the analysis of more detailed phenomena reveals the effects of the network structure.

We conclude this paper with two remarks pointing to future directions. The first one concerns the analysis of return probabilities in the Anderson model on random regular graphs put forward in \cite{Tikhonov2019}. In that paper the authors are able to write the return probability in terms of eigenfuction correlators evaluated at distance zero on the graph, which they can compute with a population dynamics method. The spatial correlation length of the eigenvector entries interests us too, in particular that of the slow decaying modes, and we aim at investigating this and related properties in the future using a similar approach. In the classical context the return probability to a node $i$ is given by $P_{ii}(t)=(e^{-\mathbf{M}t})_{ii}$, while in the quantum case one has $P_{ii}(t)=|\langle i|e^{-\mathrm{i}\mathcal{H}t}|i\rangle|^{2}$, where $\mathcal{H}$ is the Hamiltonian of the system. The representation that uses the eigenfunction correlators then holds true in our case for $(e^{-\mathbf{M}t})_{ii}^2$, which represents the probability that two independent replicas of the system starting out at the same node \emph{both} return to that node at time $t$. This quantity connects directly to our second remark, on the question of the characterization of the low temperature phase of the trap model in terms of replica symmetry breaking (RSB) in trajectory space. While the 1D lattice considered by Ueda and Sasa in \cite{Ueda2017} exhibits an RSB phase for $T<1$, this is not the case for tree-like networks, because trajectories always depart from each other in such infinite-dimensional structures. However, a possible generalization of the trajectory RSB idea to random networks is to consider closed paths only: the distribution of excursion times mentioned above exhibits a diverging mean for $T<1$, a fact that we would conjecture should be associated with an RSB phase in the space of closed trajectories. Work in this direction is in progress.

\section{Acknowledgements}

The authors acknowledge funding by the Engineering
and Physical Sciences Research Council (EPSRC)
through the Centre for Doctoral Training ``Cross Disciplinary
Approaches to Non-Equilibrium Systems''
(CANES, Grant Nr.\ EP/L015854/1). 


\bibliographystyle{unsrt} 
\bibliography{Bibliography} 

\begin{thebibliography}{10}

\bibitem{Buchner1999}
S.~B{\"{u}}chner and A.~Heuer.
\newblock {Potential energy landscape of a model glass former: thermodynamics,
  anharmonicities, and finite size effects}.
\newblock {\em Phys. Rev. E}, 60(6):6507--6518, Dec 1999.

\bibitem{Heuer2008}
A.~Heuer.
\newblock {Exploring the potential energy landscape of glass-forming systems:
  from inherent structures via metabasins to macroscopic transport}.
\newblock {\em J. Phys. Condens. Matter}, 20(37):373101, Sep 2008.

\bibitem{DeSouza2009}
V.~K. {De Souza} and D.~J. Wales.
\newblock {Connectivity in the potential energy landscape for binary
  Lennard-Jones systems}.
\newblock {\em J. Chem. Phys.}, 130(19):194508, May 2009.

\bibitem{Bray1980}
A.~J. Bray and M.~A. Moore.
\newblock Broken replica symmetry and metastable states in spin glasses.
\newblock {\em Journal of Physics C: Solid State Physics}, 13(31):907--912, Nov
  1980.

\bibitem{Cavagna1998}
A.~Cavagna, I.~Giardina, and G.~Parisi.
\newblock Stationary points of the thouless-anderson-palmer free energy.
\newblock {\em Phys. Rev. B}, 57:11251--11257, May 1998.

\bibitem{Fyodorov2012}
Y.~V. Fyodorov and C.~Nadal.
\newblock Critical behavior of the number of minima of a random landscape at
  the glass transition point and the tracy-widom distribution.
\newblock {\em Phys. Rev. Lett.}, 109:167203, Oct 2012.

\bibitem{Biroli2001}
G.~Biroli and J.~Kurchan.
\newblock Metastable states in glassy systems.
\newblock {\em Phys. Rev. E}, 64:016101, Jun 2001.

\bibitem{Berthier2011}
L.~Berthier and G.~Biroli.
\newblock {Theoretical perspective on the glass transition and amorphous
  materials}.
\newblock {\em Rev. Mod. Phys.}, 83(2):587--645, Jun 2011.

\bibitem{Chacko2019}
R.~N. Chacko, P.~Sollich, and S.~M. Fielding.
\newblock {Slow coarsening in jammed athermal soft particle suspensions}.
\newblock {\em arXiv:1903.00991}, Mar 2019.

\bibitem{Bouchaud_Cugliandolo1997}
J.~P. Bouchaud, L.~F. Cugliandolo, J.~Kurchan, and M.~M{\'{e}}zard.
\newblock {\em Out of equilibrium dynamics in spin-glasses and other glassy
  systems}, pages 161--223.
\newblock 1997.

\bibitem{Odagaki1995}
T.~Odagaki.
\newblock {Glass transition singularities}.
\newblock {\em Phys. Rev. Lett.}, 75(20):3701--3704, Nov 1995.

\bibitem{Bouchaud1997}
J.~P. Bouchaud and M.~M{\'{e}}zard.
\newblock {Universality classes for extreme-value statistics}.
\newblock {\em J. Phys. A. Math. Gen.}, 30(23):7997--8015, Dec 1997.

\bibitem{Bouchaud1992}
J.~P. Bouchaud.
\newblock {Weak ergodicity breaking and aging in disordered systems}.
\newblock {\em J. Phys. I}, 2(9):1705--1713, Sep 1992.

\bibitem{Monthus1996}
C.~Monthus and J.~P. Bouchaud.
\newblock {Models of traps and glass phenomenology}.
\newblock {\em J. Phys. A. Math. Gen.}, 29(14):3847--3869, Jul 1996.

\bibitem{Bovier2005}
A.~Bovier and A.~Faggionato.
\newblock {Spectral characterization of aging: the REM-like trap model}.
\newblock {\em Ann. Appl. Probab.}, 15(3):1997--2037, Aug 2005.

\bibitem{BenArous2006b}
G.~Ben-Arous and J.~{\v{C}ern{\'y}}.
\newblock The arcsine law as a universal aging scheme for trap models.
\newblock {\em Commun. Pure and Applied Math.}, 61(3):289--329, Dec 2006.

\bibitem{Gayrard2016}
V.~Gayrard.
\newblock {Aging in Metropolis dynamics of the REM: a proof}.
\newblock {\em HAL:01277223}, Feb 2016.
\newblock Working paper or preprint.

\bibitem{Cammarota2018}
C.~Cammarota and E.~Marinari.
\newblock Numerical evidences of universal trap-like aging dynamics.
\newblock {\em J. Stat. Mech. Theory Exp.}, 2018(4):043303, Apr 2018.

\bibitem{BaityJesi2018}
M.~Baity-Jesi, Biroli G., and Cammarota C.
\newblock Activated aging dynamics and effective trap model description in the
  random energy model.
\newblock {\em J. Stat. Mech. Theory Exp.}, 2018(1):013301, Jan 2018.

\bibitem{Ben_Arous2006}
G.~Ben~Arous and J.~{\v{C}ern{\'y}}.

\bibitem{Rinn2001}
B.~Rinn, P.~Maass, and J.~P. Bouchaud.
\newblock {Hopping in the glass configuration space: subaging and generalized
  scaling laws}.
\newblock {\em Phys. Rev. B}, 64(10):104417, Aug 2001.

\bibitem{BenArous2006}
G.~{Ben Arous}, J.~{\v{C}}ern{\'{y}}, and T.~Mountford.
\newblock {Aging in two-dimensional Bouchaud's model}.
\newblock {\em Probab. Theory Relat. Fields}, 134(1):1--43, Jan 2006.

\bibitem{Doye2002}
J.~P.~K. Doye.
\newblock {Network topology of a potential energy landscape: a static
  scale-free network}.
\newblock {\em Phys. Rev. Lett.}, 88(23):238701, May 2002.

\bibitem{Baronchelli2009}
A.~Baronchelli, A.~Barrat, and R.~Pastor-Satorras.
\newblock {Glass transition and random walks on complex energy landscapes}.
\newblock {\em Phys. Rev. E}, 80(2):020102, Aug 2009.

\bibitem{Moretti2011}
P.~Moretti, A.~Baronchelli, A.~Barrat, and R.~Pastor-Satorras.
\newblock {Complex networks and glassy dynamics: walks in the energy
  landscape}.
\newblock {\em J. Stat. Mech. Theory Exp.}, 2011(03):P03032, Mar 2011.

\bibitem{Margiotta2018}
R.~G. Margiotta, R.~K\"uhn, and P.~Sollich.
\newblock Spectral properties of the trap model on sparse networks.
\newblock {\em J. Phys. A Math. Theor.}, 51(29):294001, Jun 2018.

\bibitem{Rogers2008}
T.~Rogers, I.~P. Castillo, R.~K{\"{u}}hn, and K.~Takeda.
\newblock {Cavity approach to the spectral density of sparse symmetric random
  matrices}.
\newblock {\em Phys. Rev. E}, 78(3):031116, Sep 2008.

\bibitem{Rogers2009}
T.~Rogers and I.~P. Castillo.
\newblock {Cavity approach to the spectral density of non-Hermitian sparse
  matrices}.
\newblock {\em Phys. Rev. E}, 79(1):012101, Jan 2009.

\bibitem{Metz2010}
F.~L. Metz, I.~Neri, and D.~Boll{\'{e}}.
\newblock {Localization transition in symmetric random matrices}.
\newblock {\em Phys. Rev. E}, 82(3):031135, Sep 2010.

\bibitem{Kuhn2015}
R.~K{\"{u}}hn.
\newblock {Spectra of random stochastic matrices and relaxation in complex
  systems}.
\newblock {\em Europhys. Lett.}, 109(6):60003, Mar 2015.

\bibitem{Benedetti2018}
F.~P.~C. Benetti, G.~Parisi, F.~Pietracaprina, and G.~Sicuro.
\newblock Mean-field model for the density of states of jammed soft spheres.
\newblock {\em Phys. Rev. E}, 97(6):062157, Jun 2018.

\bibitem{Bordenave2010}
C.~Bordenave and M.~Lelarge.
\newblock {Resolvent of large random graphs}.
\newblock {\em Random Struct. Algorithms}, 37(3):332--352, Oct 2010.

\bibitem{Khorunzhy2004}
O.~Khorunzhy, M.~Shcherbina, and V.~Vengerovsky.
\newblock {Eigenvalue distribution of large weighted random graphs}.
\newblock {\em J. Math. Phys.}, 45(4):1648--1672, Apr 2004.

\bibitem{Bollobas2001}
B.~Bollobas.
\newblock {\em {Random graphs}}.
\newblock Cambridge University Press, 2001.

\bibitem{vankampen2007spp}
N.~G. Van~Kampen.
\newblock {\em Stochastic processes in physics and chemistry}.
\newblock North Holland, 2007.

\bibitem{Kurchan2009}
J.~Kurchan.
\newblock {Six out of equilibrium lectures}.
\newblock {\em arXiv:0901.1271}, Jan 2009.

\bibitem{Edwards1976}
S.~F. Edwards and R.~C. Jones.
\newblock {The eigenvalue spectrum of a large symmetric random matrix}.
\newblock {\em J. Phys. A. Math. Gen.}, 9(10):1595--1603, Oct 1976.

\bibitem{Kuhn2008}
R.~K{\"{u}}hn.
\newblock {Spectra of sparse random matrices}.
\newblock {\em J. Phys. A Math. Theor.}, 41(29):295002, Jul 2008.

\bibitem{Albert2002}
R.~Albert and A.~L. Barab{\'{a}}si.
\newblock {Statistical mechanics of complex networks}.
\newblock {\em Rev. Mod. Phys.}, 74(1):47--97, Jan 2002.

\bibitem{Mezard2001}
M.~M{\'{e}}zard and G.~Parisi.
\newblock {The Bethe lattice spin glass revisited}.
\newblock {\em Eur. Phys. J. B}, 20(2):217--233, Mar 2001.

\bibitem{Biroli1999}
G.~Biroli and R.~Monasson.
\newblock {A single defect approximation for localized states on random
  lattices}.
\newblock {\em J. Phys. A. Math. Gen.}, 32(24):L255--L261, Jun 1999.

\bibitem{Semerjian2002}
G.~Semerjian and L.~F. Cugliandolo.
\newblock {Sparse random matrices: the eigenvalue spectrum revisited}.
\newblock {\em J. Phys. A. Math. Gen.}, 35(23):303, Jun 2002.

\bibitem{Tikhonov2019}
K.~S. Tikhonov and A.~D. Mirlin.
\newblock Statistics of eigenstates near the localization transition on random
  regular graphs.
\newblock {\em Phys. Rev. B}, 99:024202, Jan 2019.

\bibitem{Ueda2017}
M.~Ueda and S.~I. Sasa.
\newblock {Replica symmetry breaking in trajectory space for the trap model}.
\newblock {\em J. Phys. A Math. Theor.}, 50(12):125001, Mar 2017.

\bibitem{Gillespie1976}
D.~T. Gillespie.
\newblock A general method for numerically simulating the stochastic time
  evolution of coupled chemical reactions.
\newblock {\em J. Comp. Phys.}, 22(4):403 -- 434, Dec 1976.

\end{thebibliography}


\appendix

%
%
\section{Mean field and random walk limits}\label{appendix_localDOS_MF_RW}
In section \ref{section_local_DOS} we saw that the power law behaviour of the local DOS close to the ground state ($\lambda=0$) arises as a combined effect of limited connectivity and trap depth disorder. When considering the two features separately, i.e.\ either taking the limit $T\to\infty$ (RW) or considering a fully connected configuration space (MF), the local DOS exhibits a $\delta$-peak at some $\lambda^*(\tau)$, and vanishes in the region $|\lambda|<|\lambda^*|$. In this appendix we discuss in more detail the local DOS of the relevant (MF and RW) limits; in particular we derive the value of $\lambda^*$ for the various cases.

\begin{figure}[htb]
	\centering
	\includegraphics{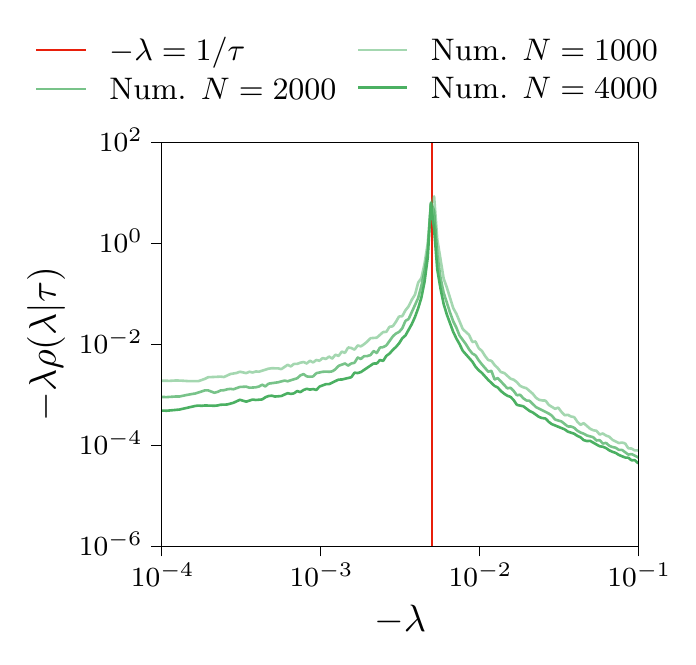}	\includegraphics{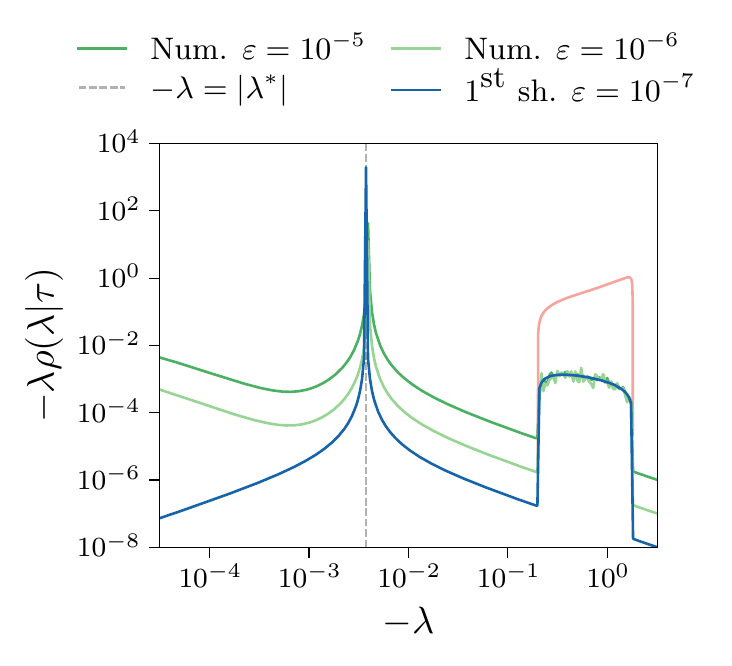}
	\caption{Left: local DOS for the mean field limit with $T=0.8$ and $\tau=200$, evaluated by averaging direct diagonalization results for systems with different size (green lines), using $\varepsilon\sim10^{-5}$. In the limit $N\to\infty$ these curves converge to a $\delta$-function centred at $\lambda=-1/\tau$ (see equation (\ref{MF_local_DOS})). Right: local DOS for the random walk limit with $c=5$ and $\tau=200$, evaluated by averaging results from direct diagonalizations of systems with size $N=4000$ (green lines), and using the exact cavity result given by (\ref{first_shell_localDOS_RW_explicit}) (blue line). The vertical dashed line indicates the location of the $\delta$-peak predicted for $\tau\gg1$ as given by (\ref{lambda_star_RRG}); the cavity results shown use non-zero $\varepsilon$ so broaden the delta peak to a narrow Lorentzian. Red curve: local DOS for $\tau=1$ with support on $\mathcal{R}$ (see main text); in this region, the cavity predictions for different epsilon and the 1st shell approximation are virtually indistinguishable.}\label{fig:MF_RW_localDOS}
\end{figure}

The easiest way to obtain a fully connected network is to consider the random $c$-regular graph ensemble and then take the limit $c\to\infty$. Imposing $p_k=\delta_{c,k}$ in (\ref{local_DOS_poulation_dynamics}) we get
\begin{equation}
\rho(\lambda|\tau) = \lim_{\varepsilon\to 0} \frac{1}{\pi} \mathrm{Re}\,\Big\langle \frac{\tau c}{\imunit\lambda_{\varepsilon}\tau c + \sum_{l=1}^{c-1}\frac{\imunit \omega_l}{\imunit + \omega_l}} \Big\rangle_{\{\omega_l\}}
\end{equation}
which for $c\gg 1$ becomes
\begin{equation}\label{local_DOS_poulation_dynamics_MF}
\rho(\lambda|\tau) = \lim_{\varepsilon\to 0} \frac{1}{\pi} \mathrm{Re}\,\Big( \frac{\tau c}{\imunit\lambda_{\varepsilon}\tau c + c\, \tilde{\omega}} \Big)
\end{equation}
where $\tilde{\omega} = \int \mathrm{d}\omega\, p(\omega) \imunit \omega/(\imunit+\omega)$. Using the self-consistency equation (\ref{cavity_precisions_distribution}) for $p(\omega)$ we get
\begin{equation}
\tilde{\omega} = \imunit \int \mathrm{d}\tau \rho_{\tau}(\tau)\, \frac{\imunit \lambda \tau c + c\, \tilde{\omega}}{\imunit + \imunit \lambda \tau c + c\, \tilde{\omega}}
\end{equation}
which implies $\tilde{\omega}=\imunit$ when $c$ is large. Substituting into (\ref{local_DOS_poulation_dynamics_MF}) then yields
\begin{equation}\label{MF_local_DOS}
\begin{split}
\rho^{\mathrm{MF}}(\lambda|\tau) &=\lim_{\varepsilon\to 0} \frac{1}{\pi}\mathrm{Re}\Big(\frac{\tau}{\varepsilon\tau + \imunit(\lambda+1)}\Big)\\
&=\lim_{\varepsilon\to 0}\frac{1}{\pi}\frac{\varepsilon}{\varepsilon^2+(\lambda+1/\tau)^2}\\
&=\delta(\lambda+1/\tau)
\end{split}
\end{equation}
It follows that $\lambda^*=-1/\tau$, and the delta peak at this location is the {\em only} contribution to the local DOS. This is displayed as a vertical line (red) in figure \ref{fig:MF_RW_localDOS}-left, which also shows the mean field local DOS evaluated by averaging results obtained from direct diagonalizations of systems with different size (shades of green). Note that finite size effects are visible in the tails away from the peak. Intuitively, the local DOS in the mean field limit has to be a $\delta$-function centred in $-1/\tau$ as the return probability and the staying probability are the same in this case, and are given by the exponential function in (\ref{staying_probability}).

The local DOS in the random walk limit can be analysed only if the value of $\tau$ is fixed \emph{before} $T$ is sent to infinity; taking $T\to\infty$ first would make all $\tau_i=1$. For the random $c$-regular graph ensemble, the local DOS is given by the approximation scheme described in section \ref{subsection_approximation_scheme} -- which is \emph{exact} in this case -- evaluated at first order. We have:
\begin{equation}\label{first_shell_localDOS_RW}
\rho^{\mathrm{1A}}(\lambda|\tau)=\lim_{\varepsilon\to 0}\frac{1}{\pi}\mathrm{Re}\Big(\frac{c\tau}{\imunit \lambda_{\varepsilon}\tau c + c \tilde{\omega}}\Big)
\end{equation}
with $\tilde{\omega}=\imunit \bar{\omega}/(\imunit+\bar{\omega})$, and $\bar{\omega}$ given by the solution of the infinite temperature cavity equation
\begin{equation}\label{omega_bar_second_orderd_eq}
\bar{\omega} = \imunit \lambda_{\varepsilon} c + (c-1)\imunit \bar{\omega}/(\imunit+\bar{\omega})
\end{equation}
The physical solution is the one with positive real part. This defines the precision of the Gaussian cavity measure (see main text before equation (\ref{precisions})) for an infinite random $c$-regular network in the limit $T\to\infty$, where the distribution of cavity precisions simplifies to $p(\omega)=\delta(\omega-\bar{\omega})$. Equation (\ref{first_shell_localDOS_RW}) can be written more conveniently as
\begin{equation}\label{first_shell_localDOS_RW_explicit}
\rho^{\mathrm{1A}}(\lambda|\tau)=\lim_{\varepsilon'\to 0}\frac{\tau}{\pi}\frac{\varepsilon'+\tilde{\omega}_{\mathrm{R}}}{(\varepsilon'+\tilde{\omega}_{\mathrm{R}})^2+(\lambda \tau +\tilde{\omega}_{\mathrm{I}})^2}
\end{equation}
where $\varepsilon'=\varepsilon \tau$ is a rescaled version of $\varepsilon$, and $\tilde{\omega}_{\mathrm{I}/\mathrm{R}}$ denote the imaginary/real part of $\tilde{\omega}$, respectively. In the limit $\varepsilon\to0$ these read
\begin{equation}
\tilde{\omega}_{\mathrm{R}}(\lambda)=\frac{\mathrm{Re}\sqrt{4c-4-c^2(1+\lambda)^2}}{2(c-1)}
\end{equation}
\begin{equation}\label{omega_tilde_im_part}
\tilde{\omega}_{\mathrm{I}}(\lambda)=\frac{c-2-c\lambda+\mathrm{Im}\sqrt{4c-4-c^2(1+\lambda)^2}}{2(c-1)}
\end{equation}
Note that $\tilde{\omega}_{\mathrm{R}}$ is positive for $\lambda \in \mathcal{R}=(-1-\Delta_c,-1+\Delta_c)$ with $\Delta_c=2(c-1)^{1/2}/c$; outside of this region it vanishes.

It is instructive to consider first the case of $\tau=1$, which gives the total DOS of the random regular graph ensemble in the random walk limit. For this simple case one can directly set $\varepsilon'=0$ to get $\rho^{\mathrm{1A}}(\lambda|1)=
\tilde{\omega}_{\mathrm{R}}/[\pi( \tilde{\omega}_{\mathrm{R}}^2+(\lambda +\tilde{\omega}_{\mathrm{I}})^2)]$, which when worked out explicitly is a scaled and shifted Kesten-McKay law \cite{Margiotta2018}.

For general $\tau>1$ one sees that the local DOS still has a contribution for $\lambda \in \mathcal{R}$ but this becomes increasingly suppressed as $\tau$ increases, scaling as $1/\tau$ for large $\tau$ from (\ref{first_shell_localDOS_RW_explicit}).
This is compensated for by an additional contribution outside of $\mathcal{R}$, where again from (\ref{first_shell_localDOS_RW_explicit})
but now with $\tilde{\omega}_{\mathrm{R}}=0$ one has $\rho^{\mathrm{1A}}(\lambda|\tau)=\delta(\lambda+\tilde{\omega}_{\mathrm{I}}(\lambda)/\tau)$, which is a delta peak at a location $\lambda^*$ determined by $-\lambda^*=
\tilde{\omega}_{\mathrm{I}}(\lambda^*)/\tau$. For large $\tau$, $\lambda^*$ becomes small so that asymptotically $\lambda^*=-\tilde{\omega}_{\mathrm{I}}(0)/\tau$, hence from  (\ref{omega_tilde_im_part}), 
\begin{equation}\label{lambda_star_RRG}
\lambda^*\simeq-\frac{c-2}{c-1}\tau^{-1}\quad \text{for}\quad \tau\gg 1
\end{equation}
This delta peak is the dominant contribution to the local DOS for large $\tau$, where using (\ref{return_probability_from_localDOS})  it gives the return probability $P_\tau(t) = \exp(-(t/\tau)(c-2)/(c-1))$ as derived by another route in (\ref{return_prob_large_tau})  in the main text. 

The right plot in figure \ref{fig:MF_RW_localDOS} shows the local DOS for the random regular graph ensemble with $c=5$, $T=0.8$ and $\tau=200$, obtained from direct diagonalizations (green lines) and using the exact cavity result (\ref{first_shell_localDOS_RW_explicit}) (blue line). The vertical dashed line corresponds to the value of $\lambda^*$ given by (\ref{lambda_star_RRG}), while the red curve shows the case of $\tau=1$ discussed above (with support only on $\mathcal{R}$). Note the strong $\varepsilon$ dependence of the results for $\lambda\neq\lambda^*$ and $\lambda\notin\mathcal{R}$, which is consistent with the expectation that $\rho(\lambda|\tau)$ vanishes in these regions in the limit $\varepsilon\to0$.

In this appendix we have only considered the case of random regular graphs, and we showed that the local DOS is composed of a continuous part with support on $\mathcal{R}$, and a $\delta$-peak whose location scales with $1/\tau$ for large $\tau$. More disordered network topologies would have the same qualitative behaviour, however. In particular, they would show a network-dependent regime of fast relaxation rates $-\lambda$, similarly to what happens in the case of finite connectivity and finite temperature discussed in section \ref{section_local_DOS} (see also \cite{Margiotta2018}-section 5 for results on the total DOS). A $\delta$-peak would again appear in the small $|\lambda|$ regime for large $\tau$.

%
%
\section{Second shell approximation}\label{appendix_second_shell_approximation}

In this appendix we derive the estimate (\ref{second_order_approximation_small_lambda}) for the power law behaviour of the local DOS close to the ground state ($\lambda=0$), which constitutes one of the central results of this work. We start by re-writing (\ref{second_order_approximation}) in a more explicit form and for a general degree distribution:
\begin{equation}\label{second_order_approximation_appendix}
\rho^{\mathrm{2A}}(\lambda|\tau) = \lim_{\varepsilon\to 0} \frac{c\tau}{\pi}\text{Re}\Big\langle \Bigl[\imunit \lambda_{\varepsilon} c \tau + \imunit k + \sum_{l=1}^{k} 1/(\imunit+f(\tau_l))\Bigr]^{-1}\Big\rangle_{\{\tau_l\},k}
\end{equation}
where 
\begin{equation}
f(\tau_l)=\imunit \lambda_{\varepsilon} c \tau_l+(c-1)\imunit \bar{\omega}/(\imunit+\bar{\omega})
\end{equation}
and $\bar{\omega}$ is the physical solution of (\ref{omega_bar_second_orderd_eq}), which for small $\lambda_{\varepsilon}$ can be approximated as
\begin{equation}
\bar{\omega}\simeq\imunit(c-1)+\imunit\frac{c(c-1)}{(c-2)}\lambda_{\varepsilon}
\end{equation}
Substituting this approximation into the definition of $f(\tau_l)$ leads to
\begin{equation}
\frac{1}{\imunit+f(\tau_l)} = \frac{\varepsilon c [\tau_l+(c-2)^{-1}] - \imunit(\lambda_{\varepsilon} c [\tau_l+(c-2)^{-1}]+c-1)}{\varepsilon^2 c^2 [\tau_l+(c-2)^{-1}]^2+ (\lambda_{\varepsilon} c [\tau_l+(c-2)^{-1}]+c-1)^2}
\end{equation}
Since we are interested in the limit $\varepsilon\to 0$ we neglect the term in the denominator multiplying $\varepsilon^2$. The error is significant only when $c^2[\tau_l+(c-2)^{-1}]^2\gg\varepsilon^{-2}$, in which case the second term of the denominator becomes very large ($\gg1$) and the resulting contribution to $\rho^{\mathrm{2A}}(\lambda|\tau)$ is negligible. We can therefore write the term in squared brackets in (\ref{second_order_approximation_appendix}) as
\begin{equation}
\begin{split}
\imunit \lambda_{\varepsilon} c \tau + \imunit k + \sum_{l=1}^{k} 1/(\imunit+f(\tau_l))
&=\varepsilon \Big[ c \tau + c \sum_{l=1}^{k}\frac{\tau_l+(c-2)^{-1}}{(\lambda c [\tau_l+(c-2)^{-1}]+c-1)^2}\Big]\\
&+\imunit\Big[\lambda c \tau +k -\sum_{l=1}^{k}\frac{1}{\lambda c [\tau_l+(c-2)^{-1}]+c-1}\Big]
\end{split}
\end{equation}
The real term on the right hand side can again be viewed as a rescaled (and still positive) $\varepsilon'$. Once this is sent to zero we obtain
\begin{equation}\label{second_order_approximation_appendix_yl}
\rho^{\mathrm{2A}}(\lambda|\tau) = c\tau\Big\langle \delta(\lambda c \tau +k -\sum_{l=1}^{k} y_l)\Big\rangle_{\{\tau_l\},k}
\end{equation}
with
\begin{equation}
y_l = \frac{1}{\lambda c [\tau_l+(c-2)^{-1}]+c-1}
\end{equation}
Since we are interested in the regime $|\lambda|\ll1/\tau$, the term $\lambda c \tau$ in the $\delta$-function of equation (\ref{second_order_approximation_appendix_yl}) can be discarded. We observe then that the only non-zero contributions to the local DOS are given by those combinations of $\{\tau_l\}$ that result in $\sum_l y_l=k$. To understand when this happens let us set
\begin{equation}
a = c-1 -|\lambda| c/(c-2), \qquad b = |\lambda|c
\end{equation}
so that $y_l=(a-b\tau_l)^{-1}$. From the distribution of lifetimes (\ref{lifetime_distribution}) we obtain
\begin{equation}
\rho_y(y_l) = T b^T(a-1/y_l)^{-(T+1)}y_l^{-2}
\end{equation}
for $y_l<0$ or $y_l>1/(a-b)$. Note that as $|\lambda|\to0$, also $b\to 0$. In this limit, $\rho_y(y_l)$ goes to zero everywhere except in a region of order $b$ around $y_l = 1/(c-1)$. We can now approximate the probability distribution of $Y=\sum_{l=1}^{k}y_l$. This drops by a factor $b^T$ for each of the $y_l$ that is away of $1/(c-1)$, so the most likely way to realize $Y=k$ is to have $k-1$ of the $y_l$ equal to $1/(c-1)$, and only a single one, say $y_1$, equal to $k-(k-1)/(c-1)$. Note that this happens when $\tau_1\sim 1/|\lambda|$, i.e.\ if there is a single deep minimum in the first neighbouring shell of the departing trap with lifetime as large as $1/|\lambda|$. So we have $p(Y=k)\simeq k \rho_y(k-(k-1)/(c-1))
$, where the factor $k$ arises because any of the $y_l$ could be the large one. Finally, from (\ref{second_order_approximation_appendix_yl}) (with $\lambda c \tau \to 0$) we see that
\begin{equation}
\begin{split}
\rho^{\mathrm{2A}}(\lambda|\tau) &\simeq c\tau \Big\langle \delta(k-Y) \Big\rangle_{Y,k}\\
& \simeq c\tau \Big\langle k \rho_y(k-(k-1)/(c-1))\Big\rangle_{k}\\
& = \tau T \bar{\alpha}_k c^{T+1} (c-1)^{1-T} |\lambda|^T
\end{split}
\end{equation}
which is the same as equation (\ref{second_order_approximation_small_lambda}) in the main text, with $C(c,T) = T \bar{\alpha}_k c^{T+1}(c-1)^{1-T}$ and
\begin{equation}
\bar{\alpha}_k = \Big\langle k((c-2)k+1)^{-2} \Big(1-\frac{c-1}{(c-2)k+1}\Big)^{-(T+1)} \Big\rangle_k
\end{equation}
The simplest case of the random regular graph ensemble is obtained by imposing $p_k=\delta_{c,k}$ in the last equation, which leads to $C(c,T)=Tc(c-1)^{T-1}(c-2)^{-(T+1)}$.

%
%
\section{Simulated dynamics}\label{appendix_simulations}
The data labelled ``simulations" shown in the figures \ref{fig:p_vs_t} and \ref{fig:P_F_S}-right have been collected by using a version of the \emph{stochastic simulation algorithm} (SSA) that generates dynamically an infinite tree. The SSA became popular after Gillespie applied it to the study of chemical reactions \cite{Gillespie1976}, and for this reason it is also known as ``Gillespie algorithm". Its general idea is to implement the stochastic evolution of a system on a discrete state-space as follows: for each state $i$, a random waiting time $\mathrm{d}t_i$ is sampled from the exponential distribution $r_i^{\mathrm{ex}}\exp(-r_i^{\mathrm{ex}} t)\theta(t)$, where $r_i^{\mathrm{ex}}=\sum_{j(\neq i)} r_{ji}$ is the total exit rate from state $i$, and $r_{ji}$ is the transition rate from state $i$ to state $j$. Then, the next state $j$ is chosen with probability $r_{ji}/r_i^{\mathrm{ex}}$. This process is repeated until the total time $\sum_i \mathrm{d}t_i$ exceeds some $t_{\mathrm{max}}$ that sets the maximum running time of the simulation. 

In our case, the states are represented by the nodes of the network. These have four main attributes that we track in the simulation: the energy $E_i$, the degree $k_i$, the distance (from the origin) $d_i$ and the (number of) visits $n_i$. The time spent in a given node $i$ and the next node visited $j$ are defined by the routine \emph{gillespietrap}$(i)$. This works as follows:
\begin{enumerate}
	\item compute the total exit rate $r_i^{\mathrm{ex}}=\sum_{j\in\partial i}r_{ji}=k_i e^{-\beta E_i}/c$ ($\partial i$ indicates the neighbourhood of node $i$);
	
	\item compute the waiting time $\mathrm{d}t$  by sampling from $p_i(t)=r_i^{\mathrm{ex}}\exp(-r_i^{\mathrm{ex}} t)\theta(t)$;
	
	\item select the next node $i^{\mathrm{new}}$ randomly from the $k_i$ neighbours: $r_{ji}/r_i^{\mathrm{ex}}=1/k_i$;
	
	\item return $i^{\mathrm{new}}$ and $\mathrm{d}t$.
\end{enumerate}
The quantities of interest, such as the current state or the distance from the origin, are measured at times defined by a time-grid with $n_{\mathrm{times}}$ values in the range $[0,t_{\mathrm{max}}]$. In order to simulate the evolution on an infinite tree, the algorithm has to create the network structure on the fly. This can be done as follows:
\begin{enumerate}
	\item start from a node with $k_0$ leaves ($k_0$ is sampled from $p_k$), energy $E_0$, distance $d_0=0$ and visits $n_0=1$. Each leaf $j\in \partial i$ has a random energy sampled from $\rho_{E}(E)$, degree $k_j=1$, distance $d_j=1$ and visits $n_j=0$. This is the starting network configuration;
	
	\item select the next node $j$ with \emph{gillespietrap}$(i)$. If $n_j=0$, attach a new neighbourhood to $j$, taking into account that $j$ already has $i$ as neighbour. This is done by the routine \emph{newneighbourhood}$(j)$. Then, the number of visits is set to $n_j=n_j+1$;
	
	\item \emph{newneighbourhood}$(j)$ assigns $k^{\mathrm{new}}_j-1$ leaves to $j$, with $k^{\mathrm{new}}_j$ sampled from $kp_k/c$, and so it replaces $k_j=1$ with $k^{\mathrm{new}}_j$. The new leaves $l\in\partial j \setminus i$ have random energies $E_l$ sampled from $\rho_{E}(E)$, distance $d_l=d_j+1$, degree $k_l=1$ and visits $n_l=0$.
\end{enumerate}
Note that no loops, single nodes or disconnected components are created. The algorithm can also be used for running multiple copies of the dynamics in parallel, which is useful if one is interested in collecting data for a given realization of the disorder. The full structure of the implementation is explained in the following pseudo-code, where $x(m)$ indicates the position of the $m^{\mathrm{th}}$ copy of the system, $t(m)$ its time, and there are $n_{\mathrm{copies}}$ in total.\\[0.25cm]
\newpage
\noindent\textbf{Pseudo-code:} simulations on the infinite tree.\\[0.25cm]
\begin{algorithmic}
	\State set all $x(m)=0$ {\color{mygrey}*where $0$ is the initial trap*}
	\State set all $t(m)=0$
	\State set all $x^{\mathrm{new}}(m)=x(m)$
	\State create the starting network configuration
	\State $\ldots$
	\For{($t=0$; $t< n_{\mathrm{times}}$; $t$++)}
	\State $\mathrm{time}=\mathrm{time\_grid}(t)$
	\For{($m=1$; $m\leq n_{\mathrm{copies}}$; $m$++)}
	\While{($t(m)\leq \mathrm{time}$)}
	\State {\color{mygrey}*carry out one transition*}
	\State $x(m)= x^{\mathrm{new}}(m)$
	\State $x^{\mathrm{new}}(m),\,\mathrm{d}t=\mathrm{\emph{gillespietrap}}(x(m))$
	\If{($(x^{\mathrm{new}}(m)).\mathrm{\emph{visits}}==0$)} 
	\State \emph{newneighbourhood}$(x^{\mathrm{new}}(m))$
	\EndIf
	\State $t(m)=t(m)+\mathrm{d}t$
	\EndWhile
	\State {\color{mygrey}*collect statistics here*}
	\State $\ldots$
	\EndFor
	\EndFor
\end{algorithmic}

\end{document}